\documentclass[10pt,journal,compsoc]{IEEEtran}
\usepackage{algorithmic}
\usepackage{algorithm}
\usepackage{amsmath,amsthm,amsfonts}
\usepackage{graphicx}
\usepackage{subfig}
\usepackage{bbm}
\usepackage{bm}
\usepackage{overpic}
\usepackage{epstopdf}
\usepackage{tikz}
\usepackage{overpic} 
\usepackage{url}
\usepackage{verbatim} % used for comment environment
\usepackage{amssymb} % used for \intercal (transpose)
\usepackage{bbm}

\usetikzlibrary{automata,arrows,positioning,calc}
\definecolor{mycolor}{RGB}{91,155,213}

\newtheorem{lemma}{Lemma}
\newtheorem{theorem}{Theorem}
\newtheorem{definition}{Definition}
\newtheorem{remark}{Remark}
\newtheorem{corollary}{Corollary}

\ifCLASSOPTIONcompsoc
  \usepackage[nocompress]{cite}
\else
  \usepackage{cite}
\fi
  
\begin{document}
\title{Optimal Routing for Delay-Sensitive Traffic in Overlay Networks}
\author{Rahul~Singh,~\IEEEmembership{Member,~IEEE,}
and~Eytan~Modiano,~\IEEEmembership{Fellow,~IEEE,} %
\IEEEcompsocitemizethanks{\IEEEcompsocthanksitem Rahul Singh and Eytan Modiano are with the Laboratory for Information \& Decision Systems (LIDS), Massachusetts Institute of Technology, Cambridge, MA 02139, USA.\protect\\
E-mail: rahulsiitk@gmail.com, modiano@mit.edu.}% <-this % stops an unwanted space
\thanks{This work was supported by NSF grant CNS-1524317, and by DARPA I2O and Raytheon BBN Technologies  under Contract No. HROO l l-l 5-C-0097.}    
}
\IEEEtitleabstractindextext{
\begin{abstract}\label{sec:abs}
We design dynamic routing policies for an overlay network which meet delay requirements of real-time traffic being served on top of an underlying legacy network, where the overlay nodes do not know the underlay characteristics. We pose the problem as a constrained MDP, and show that when the underlay implements static policies such as FIFO with randomized routing, then a decentralized policy, that can be computed efficiently in a distributed fashion, is optimal. Our algorithm utilizes multi-timescale stochastic approximation techniques, and its convergence relies on the fact that the recursions asymptotically track a nonlinear differential equation, namely the replicator equation. Extensive simulations show that the proposed policy indeed outperforms the existing policies.       
\end{abstract}
}
\maketitle
\IEEEdisplaynontitleabstractindextext

%\IEEEpeerreviewmaketitle

\section{Introduction}\label{sec:intro}
\begin{figure}
\includegraphics[scale=.28]{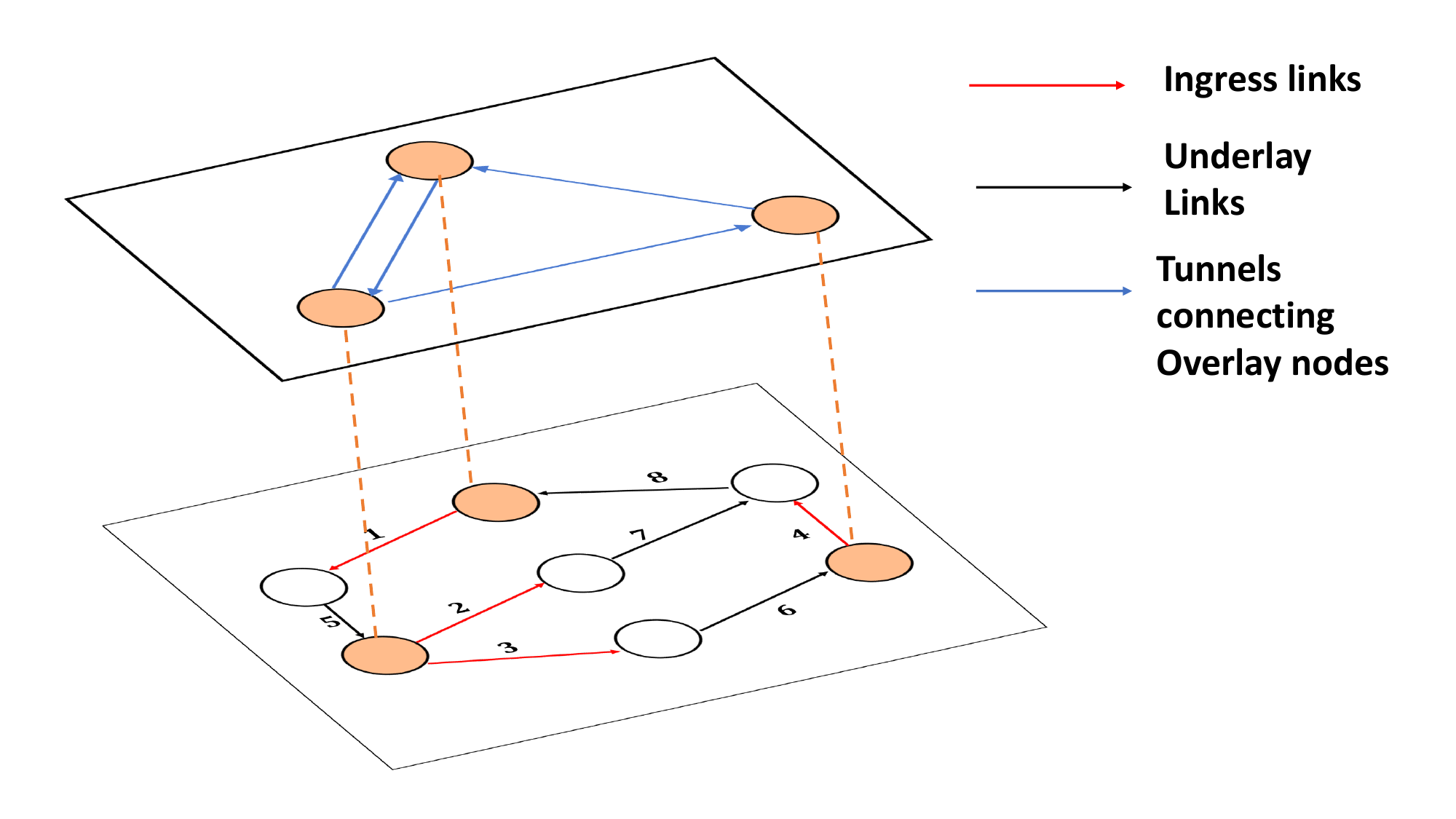}
%\vspace{-.5in}
\caption{The bottom plane shows the complete network in which the Overlay nodes are colored. The top plane visualizes the overlay network, in which a tunnel corresponds to an existing, possibly randomized underlay path. The ingress links $1,2,3,4$ inject traffic from overlay nodes into the underlay nodes.}
\label{f1}
\end{figure}

Overlay networks are a novel concept to bridge the gap between what modern Internet-based services need and what the existing networks actually provide, and hence overcome the shortcomings of the Internet architecture~\cite{sitaraman2014overlay,peterson2007computer}. The overlay creates a virtual network over the existing underlay, utilizes the functional primitives of the underlay, and supports the requirements of modern Internet-based services which the underlay is not able to do on its own (see Fig.~\ref{f1}). The overlay approach enables new services at incremental deployment cost.

The focus of this paper is on developing efficient overlay routing algorithms for data which is generated in real-time, e.g., Internet-based applications such as banking, gaming,  shopping, or live streaming. Such applications are sensitive to the end-to-end delays experienced by the data packets.

Dynamic routing policies for multihop networks have been traditionally studied in the context where all nodes are controllable~\cite{tassi1,sasha1}. Our setup, however, allows only a subset of the network nodes (overlay) to make dynamic routing decisions, while the nodes of the \emph{legacy network} (underlay) implement simple policies such as FIFO combined with fixed path routing. This approach introduces new challenges because the overlay nodes do not have knowledge of the underlay's topology, routing scheme or link capacities. Thus, the overlay nodes have to \emph{learn} the optimal routing policy in an \emph{online} fashion which involves an exploration-exploitation trade-off between finding low delay paths and utilizing them. Moreover, since the network conditions and traffic demands may be time-varying, this also involves consistently ``tracking" the optimal policy.

 \subsection{Previous Works} %\vspace{-.25in}
The use of overlay architecture was originally proposed in~\cite{andersen2001resilient} to achieve network resilience by finding new paths in the event of outages in the underlay. In a related work~\cite{han2005topology} considered the problem of placing the underlay nodes in an optimal fashion in order to attain the maximum ``path diversity". In~\cite{jones2014overlay} the authors consider optimal overlay node placement to maximize throughput, while~\cite{paschos2014throughput} develops throughput optimal overlay routing in a restricted setting.

While many works~\cite{boyan1994packet,awerbuch2004adaptive} use end-to-end feedbacks for delay optimal routing, these works ignore the \emph{queueing aspect}, and hence the delay of a packet is assumed to be independent of the congestion in the underlay network. 
Early works on delay minimization~\cite{gallager1977minimum,tsitsiklis1986distributed} concentrated on quasi-static routing, and do not take the network state into account while making routing decisions.
The existing results on dynamic routing explicitly assume that all the network nodes are controllable, and typically analyze the performance of algorithms when the network is heavily loaded~\cite{sasha1}. However, in the heavy traffic regime, the delays incurred under any policy are necessarily large, and thus not suitable for routing of real-time traffic. Finally, we note that the popular backpressure algorithm is known to perform poorly with respect to average delays~\cite{bui2009novel}.    
\subsection{Contributions}
In contrast to the approaches mentioned above, we consider a network where only a subset of nodes (overlay) are controllable, and propose algorithms that meet average end-to-end delay requirements imposed by applications. Our algorithms are decentralized and perform optimally irrespective of the network load. 

It follows from Little's law~\cite{soren} that for a stable network, the objective of meeting an average end-to-end delay requirement can be replaced by keeping the average queue lengths below some value $B$. In this work, the problem of maintaining the average queue lengths below $B$ is divided into two sub-problems: i) distributing the bound $B$ into link-level average queue thresholds $B_\ell$ such that these link-level bounds can be satisfied under some policy, ii) designing an overlay routing policy that meets the link-level queue bounds imposed by i).

We obtain an efficient decentralized solution to ii) by introducing the notion of ``link prices" that are charged by links. 
The link prices induce cooperation amongst the overlay nodes, thereby producing decentralized optimal routing policy, and are in spirit with the Kelly decomposition method for network utility maximization~\cite{kelly}. The average queue lengths are adaptively controlled~\cite{astrom,borkaradapt} by manipulating the link prices, but unlike previous works which utilize Kelly decomposition in a static deterministic setting~\cite{palomar2006tutorial}, we perform a stochastic dynamic optimization with respect to the routing decisions.

\begin{figure}
\includegraphics[scale=.4]{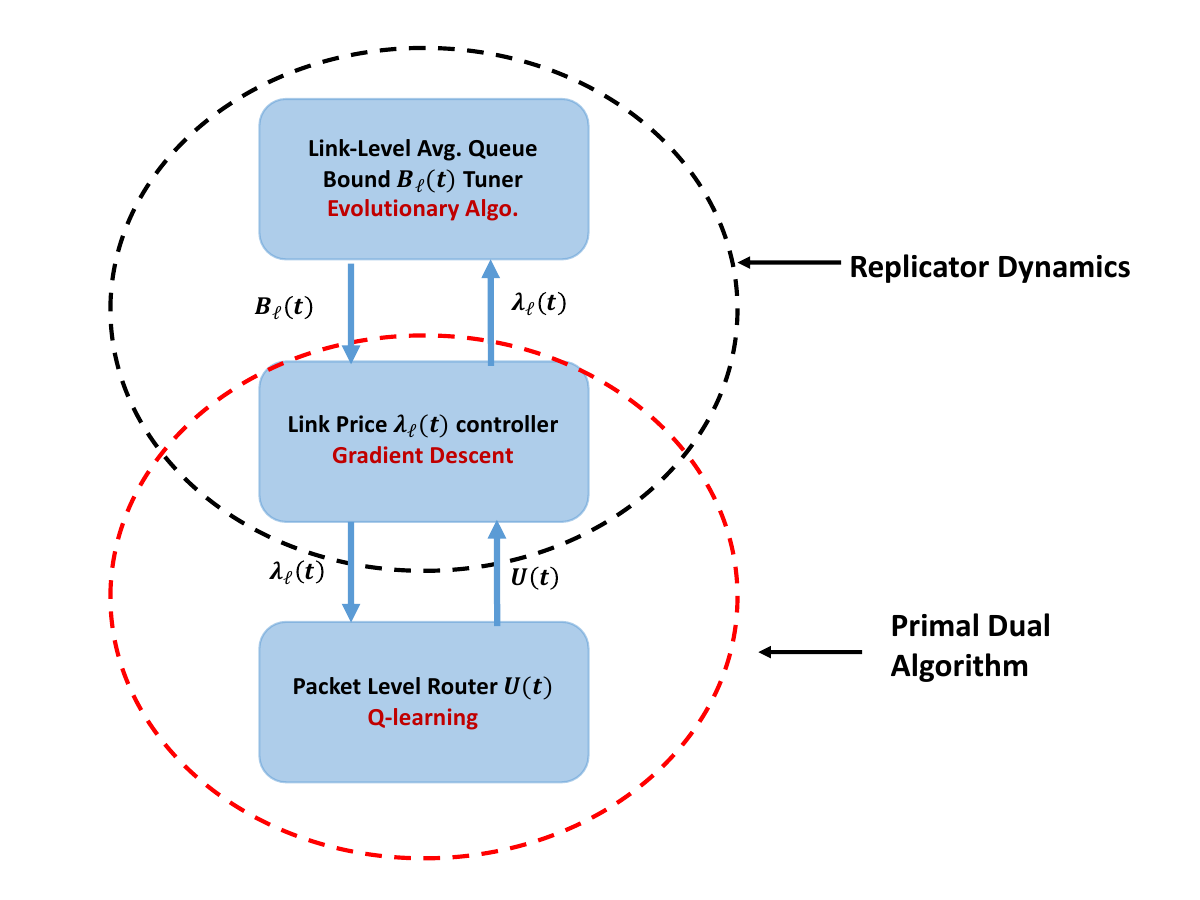}
%\vspace{-.5in}
\caption{The proposed $3$ layer Price-based Overlay Controller (POC) comprising of (from bottom to top) i) Overlay nodes making packet-level routing decisions $U(t)$,
ii) link-level price controller which manipulates the prices $\lambda_\ell(t)$, iii) Link-level average queue threshold  manipulator which tunes the $B_\ell(t)$. The interactions between the top 2 layers are described by a nonlinear ode called replicator dynamics, while the bottom 2 layers constitute a primal-dual algorithm.}
\label{f3}
\end{figure}

\begin{figure}
\includegraphics[scale=.25]{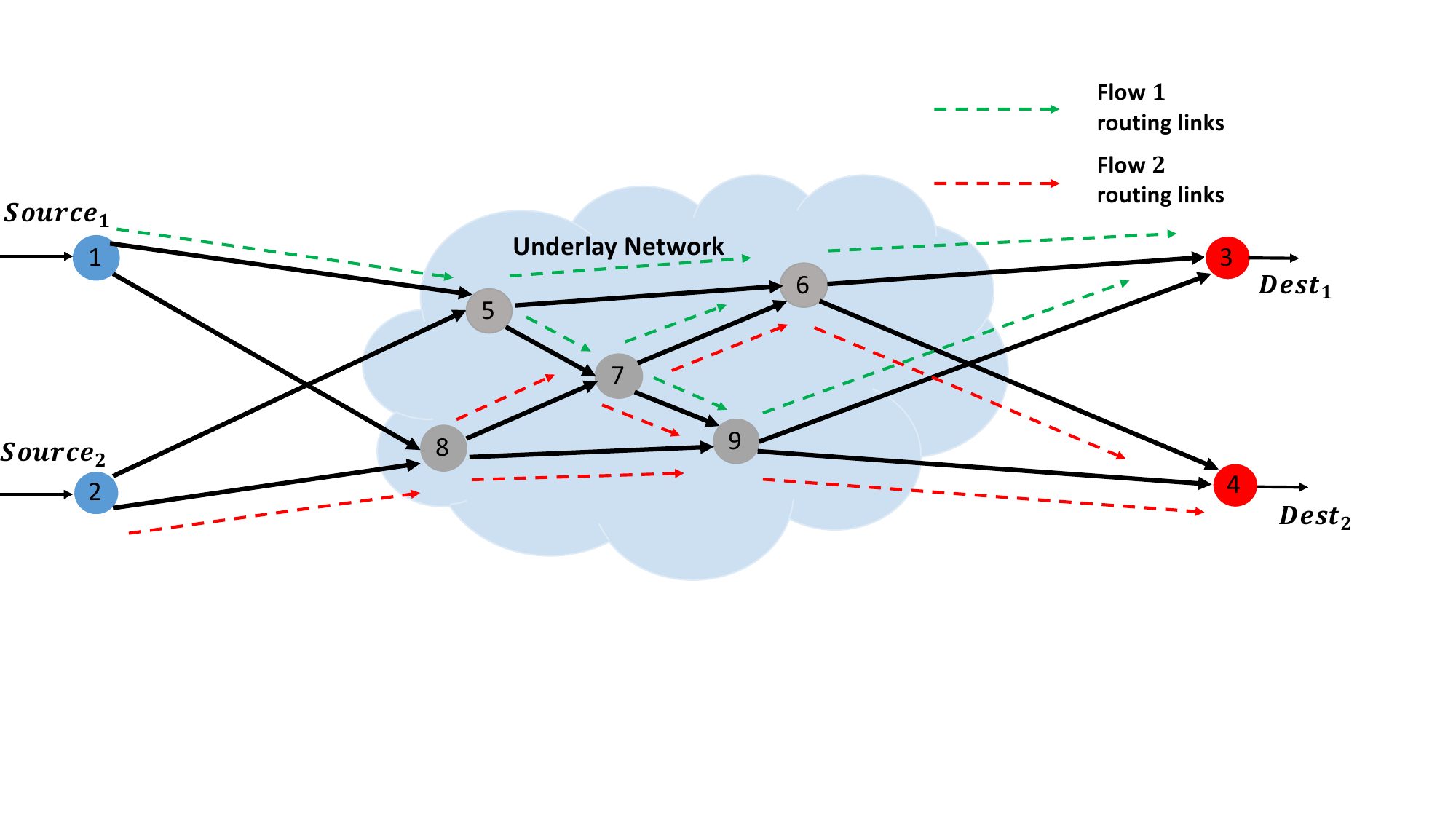}
\vspace{-.55in}
\caption{An overlay network comprising of $2$ source destination pairs $(1,3)$ and $(2,4)$. Since the routes taken by packets sent at ingress links $(1,5)$ and $(2,8)$ share common underlay links $(7,9)$ and $(7,6)$, an increase in traffic intensity sent on either of the ingress links may also increase the delay suffered by packets sent on the other ingress link.}
\label{f2}
\end{figure}

In order to solve i) we provide an adaptive scheme which follows the replicator dynamics~\cite{weibull1995evolutionary}. Finally, the solutions to i) and ii) are combined to yield a 3 layer queue control overlay routing scheme, see Figs.~\ref{f3} and~\ref{f4}.
Our problem also has close connections to the restless Multi Armed Bandit problem (MABP)~\cite{Whittle2011Book,lai1985asymptotically}. Our scheme takes an ``explore-exploit strategy" in absence of knowledge regarding the underlay network's characteristics, and learns the optimal routing policy in an online fashion using the data obtained from network operation. The routing decisions on the various ingress links (see Fig~\ref{f2}) correspond to the bandit ``arms", while the average end-to-end network delays are the \emph{unknown} rewards. Since the packets injected by different ingress links share common underlay links on their path, routing decisions at an ingress link $\ell$ affect the delays incurred by packets sent on different ingress link $\hat{\ell}$ (see Fig.~\ref{f2}). This introduces dependencies amongst the Bandit arms, and hence the decision space grows exponentially with the number of ingress overlay links. Consequently we cannot apply existing MABP algorithms, and must develop simpler algorithms that suit our needs. Furthermore, we also notice that the delay induced on each link $\ell$ of the network is also a function of the routing decisions taken at the source nodes. Hence, we cannot use the existing results from combinatorial multi armed bandit literature such as~\cite{gai2012combinatorial,chen2013combinatorial,kveton2015tight,he2013endhost} which assume that the probability distribution of the ``reward"  yielded by the arms does not depend upon the choice of the arms played. This assumption which is not true for our setup, since the distribution of the reward (delay) is also a function of the routing decisions (arms chosen to be played).

Our goal is to develop decentralized policies to control the end-to-end delays, which can be computed efficiently in a parallel and distributed fashion~\cite{bertsekas1989parallel}. We show that if the underlay implements a simple static policy, then there exists a decentralized policy that is optimal. When the underlay is allowed to use dynamic policies, we provide theoretical guarantees for our decentralized policies.

We begin in Section~\ref{sec:pf} by describing the set-up, and pose the problem of designing an overlay policy to keep the average end-to-end delays within a pre-specified bound, as a constrained Markov Decision Process (CMDP)~\cite{altman}. Section~\ref{sec:fld} solves the problem of meeting link-level average queue bounds.
It is shown that the routing decisions across the flows can be decoupled if the links are allowed to charge prices for their usage. This \emph{flow-level decomposition} technique significantly simplifies the policy design procedure, and also ensures that the resulting scheme is decentralized. Section~\ref{sec:atd} employs an evolutionary algorithm to tune the link-level average queue bounds, and proves the convergence properties. Section~\ref{sec:ext} discusses several useful extensions. Section~\ref{sec:simu} compares the performance of our proposed algorithms with the existing algorithms.  %We finally conclude in Section~\ref{sec:con}.
\section{Problem Formulation}\label{sec:pf}
We will first describe the system model, and then proceed to pose the problem of bounding the average end-to-end delays.
\subsection{System Model}
The network is represented as a graph $G=\left(N,E\right)$, where $N$ is the set of nodes, and $E$ is the set of links. The network evolves over discrete time-slots. A link $\ell = \left(i,j\right)\in E$ with capacity $C_l(t), t=1,2,\ldots$ implies that node $i$ can transmit $C_l$ packets to node $j$ at time $t$. We allow for the link capacities $C_\ell(t)$ to be stochastic, i.e., $C_\ell(t)$ depends on the state of link $\ell$ at time $t$. We will assume that the link states are i.i.d. across time \footnote{Our analysis extends in a straightforward manner for the case when the states evolve as a finite state Markov process.}.     

Multiple flows $f=1,2,\ldots,F$ share the network. Each flow $f$ will be associated with a source node $s_f$ and destination node $d_f$. We will assume that the packet arrivals at each source node $s_f$ are i.i.d. across time and flows\footnote{Our analysis can easily be extended for the case when arrivals are governed by a finite state Markov process.}. Mean arrival rate at source node $s_f$ will be denoted by $A_f$. The number of arrivals at any source node are uniformly bounded across time and flows. 

There are two types of nodes in $G$: i) \emph{overlay}: Those that can make \emph{dynamic} routing and scheduling decisions based on the network state, ii) \emph{underlay}: Those that  implement FIFO scheduling combined with randomized routing, on a per flow basis.

The subgraph induced by the underlay nodes will be called the underlay network or just underlay. In order to make the exposition simpler and avoid cumbersome notation, we will make some simplifying assumptions\footnote{Our algorithms, and their analysis can be easily extended to the case where these assumptions are not satisfied. We choose to present the simple case in order to simplify the exposition of ideas, and avoid unnecessary notation.}. Firstly, the overlay will be assumed to be composed entirely of source and destination nodes. Thus the set of overlay nodes is given by,
\begin{align*}
\{i\in N:i=s_f\mbox{ or } i=d_f \mbox{ for some flow } f=1,2,\ldots,F \}.
\end{align*}
Under this assumption, there are no multiple alternating segments of overlay-underlay nodes connecting the source-destination pairs. We will assume that the overlay nodes are connected in the overlay network only through underlay tunnels (see Fig.~\ref{f1}), i.e., there are no ``direct links" of the type $\left(s_f,d_f\right)$ that connect the source and destination links. Also, the flows do not share source nodes. % Finally, we do not allow queueing at the source nodes $s_f$, i.e., at each time $t$, each source node $s_f$ forwards the packets that arrive to it at time $t$ at various ingress links.

\emph{Ingress Links} : The network links $\{\ell=(i,j): i\in \mbox{Overlay}, j\in \mbox{Underlay}\}$ will be referred to as the ingress links. These are the outgoing links from overlay nodes that connect with the underlay, and hence these are precisely the links where the overlay routing decision take place (see Fig.~\ref{f1}). 

\begin{figure}
\includegraphics[scale=.30]{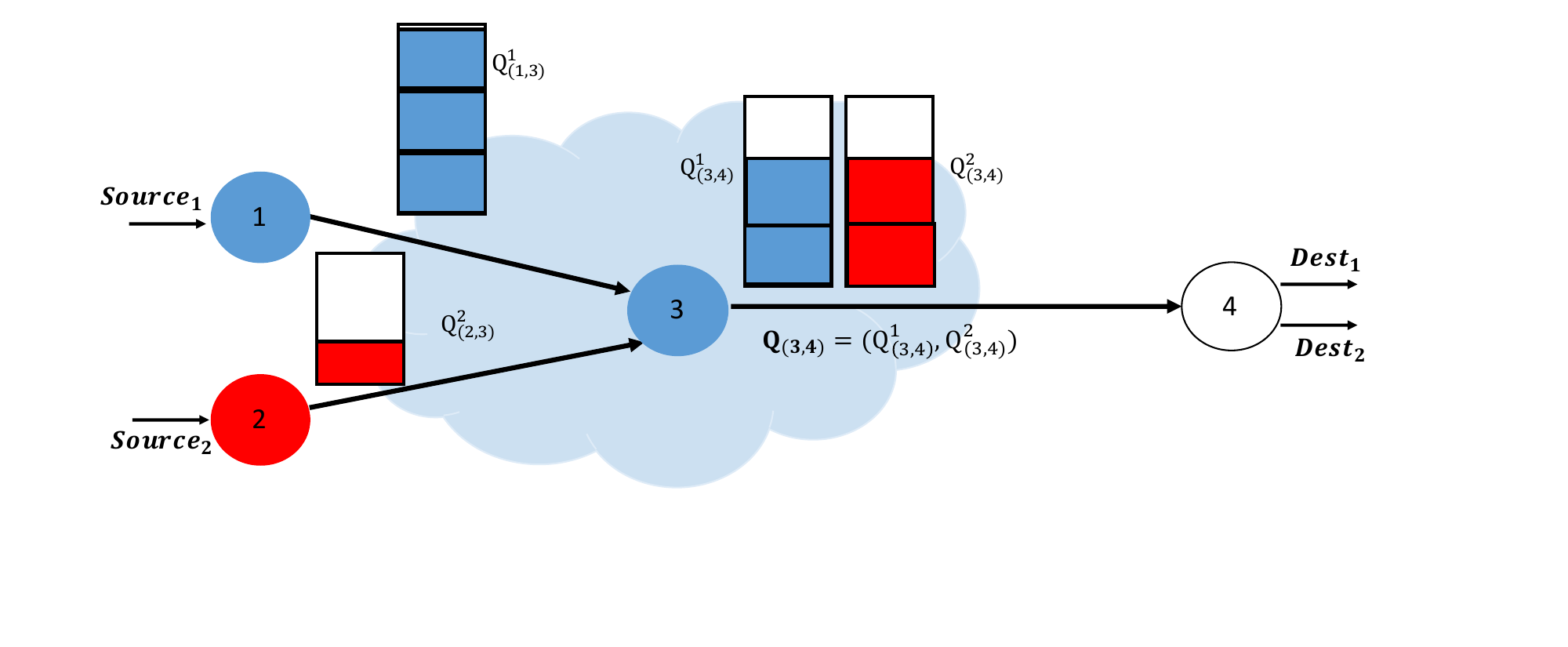}
%\vspace{-.70in}
\caption{An overlay network comprising of $2$ flows that share the same destination node $4$. Each link $\ell$ maintains separate queues for the flows $f$ that are routed on it. Queues on link $(3,4)$ are given by the vector $\bm{Q}_{(3,4)} = \left(Q^1_{(3,4)},Q^2_{(3,4)}\right)$.
}
\label{f5}
\end{figure}

\emph{Underlay Operation} : The network evolves over discrete time slots indexed $t=0,1,\ldots$. Each \emph{underlay link} $\ell$ maintains a separate queueing buffer for each flow in which it stores packets belonging to that flow (see Fig.~\ref{f5}). An underlay link $\ell$ is shared by queues belonging to flows whose routes utilize link $\ell$. 
In order to simplify the exposition of ideas, we begin by assuming that each underlay link $\ell$ implements a static scheduling policy by sharing the available capacity at any time $t$ amongst the flows in some pre-decided ratio. Thus, there are constants $\mu_{\ell,f}, f=1,2,\ldots,F\quad \ell\in E$ satisfying $\mu_{\ell,f}\geq 0, \sum_f\mu_{\ell,f}=1$. At each time $t$, flow $f$ receives a proportion $\mu_{\ell,f}$ of the available link capacity at link $\ell$. Note that in case a flow $f$ does not use the link $\ell$ for routing, then $\mu_{\ell,f}=0$. The links can use randomization in order to allocate the capacity in the ratio $\mu_{\ell,f}$. 

The underlay employs a randomized routing discipline. Thus, once a flow $f$ packet is successfully transmitted on an underlay link $\ell$, it is routed with a probability $P^f(\ell,\hat{\ell})$ to link $\hat{\ell}$. Note that $\sum_{\hat{\ell}}P^f(\ell,\hat{\ell})=1\quad\forall f,\ell$, and the routing probabilities are allowed to be flow-dependent. We denote by $R_f$, the set of links $\ell$ which are used for routing flow $f$ packets. Note that this includes static routing such as shortest path.

Our analysis extends easily to include the case where the underlay can utilize dynamic routing-scheduling disciplines such as longest queue first, priority based scheduling, or even the Backpressure policy~\cite{tassi1}. We discuss this extension in Section~\ref{sec:ext}.
% of our technical report~\cite{dropbox}. %Link capacities $C_\ell$ denote the maximum number of packets that a link $\ell$ can transmit in any time-slot. 

\emph{Information available to the Overlay}: Overlay nodes do not know the underlay's topology, probability laws governing the random link capacities, nor the policy implemented by it. In Sections~\ref{sec:fld}, and~\ref{sec:atd} we will assume that each source node $s_f$ knows the underlay queue lengths corresponding to its flow but not the overall underlay queue length or the queue lengths corresponding to other flows. Later, we will relax this assumption and assume that source nodes know only the end-to-end delays corresponding to its flows.

\emph{Actions available to the Overlay}: Let $\bm{Q}(t):=\left(\bm{Q}_1(t),\bm{Q}_2(t),\ldots,\bm{Q}_L(t)\right)$, where each $\bm{Q}_\ell(t)$ is a vector containing queue lengths $Q^f_\ell(t)$ of all the flows $f$ that are routed through link $\ell$ (see Fig.~\ref{f5}). At each time $t\geq 0$, each overlay source node $s_f$ decides its action $\bm{U}^f(t):=\{U^f_\ell(t)\}_{\ell=(s_f,i)}$,where $U^f_\ell(t)$ is the number of flow $f$ packets sent to the queueing buffer of the ingress link $\ell$ at time $t$. This decision is made based on the information available to it until time $t$ which includes $\{\bm{U}^f(s),\bm{Q}^f(s),\bm{\lambda}(s)\}_{s=0}^{t-1}$, where $\bm{Q}^f(t)$ is the vector of queue lengths corresponding to the flow $f$, and $\bm{\lambda}(s)=\{\lambda_\ell(s)\}_{\ell\in E}$ is the vector comprising of link prices at time $s$.
\\
% xxx \textbf{IC 12: define policy here }\\

\emph{Scheduling Policy}:  We will denote vectors in boldface. A routing policy $\pi$ is a map from $\bm{Q}(t)$, the vector of queue lengths at time $t$, to a scheduling action $\bm{U}(t)=\{\bm{U}^f(t)\}_{f=1,2,\ldots,F}$ at time $t$. The action $\bm{U}^f(t)$ determines the number of packets to be routed at time $t$ on each of the ingress links belonging to flow $f$.

\emph{Policy at Underlay } Throughout, we will assume that the underlay links implement a simple static policy in which each link $\ell$ shares the link capacity $C_\ell$ amongst the flows in some constant ratio, even if some of the queues $Q^f_\ell(t)$ are empty. In Section~\ref{sec:ext} we will provide guidelines to consider the case when such an assumption is not true, i.e., the underlay network implements a more complex policy.
\subsection{Objective: Keep Average Delays Bounded by $B$}
We now pose the problem of designing an overlay routing policy that meets a specified requirement on the average end-to-end delays incurred by packets as a constrained MDP~\cite{altman}.

For a stable network, it follows from the Little's law~\cite{soren}, that for a fixed value of the mean arrival rate, the mean delay is proportional to the average queue lengths. Thus, we will focus on controlling the average queue lengths instead of bounding the end-to-end delays. A bound on the average queue lengths will imply a bound on the average end-to-end delays.

Since now our objective is to control the average queue lengths, the state \cite{bertsekasdp} of the network at time $t$ is specified by the vector $\bm{Q}(t)$. Also, we let $\|\bm{Q}(t)\|:=\sum_{\ell,f} Q^f_\ell(t)$ denote the total queue length, i.e. the $1$ norm of the vector $\bm{Q}(t)$. We also let $\bm{Q}^f(t):= \{Q^f_\ell(t)\}_{\ell\in R_f}$ denote the vector containing queue lengths belonging to flow $f$. The queue dynamics are given by,
\begin{align*}
Q^f_\ell(t+1) = \left(Q^f_\ell(t)-D^f_\ell(t)\right)^{+} + A^f_\ell(t),\forall f,\ell,
\end{align*}
where $D^f_\ell(t),A^f_\ell(t)$ are the number of flow $f$ departures and arrivals respectively at link $\ell$ at time $t$. The arrivals could be external, or due to routing after service completions at other links.

If the overlay knew the underlay topology, link capacities and routing policy, the overlay could solve the following constrained MDP~\cite{altman} in order to keep the average queue lengths less than the threshold $B$,
\begin{align}
&\qquad \min_\pi 0\label{cmdp:del}\\
& \lim_{T\to\infty}\frac{1}{T} \mathbb{E}_\pi\left\{\sum_{t=1}^{T}\|\bm{Q}(t)\|\right\}\leq B,\label{cmdp:del1}
\end{align} 
where expectation is taking with respect to the (possibly random) routing policies at the overlay and underlay, and the randomness of arrivals and link capacities. 
If we assume that the network queue lengths are uniformly bounded, i.e., $Q^f_\ell(t)\leq Q_{\max},\quad \forall \ell \in E, f=1,2,\ldots,F$, then standard theory from constrained MDP~\cite{altman} tells us that the above is solved by a stationary randomized policy. Thus $\bm{U}(t)=\{\bm{U}^f(t)\}_{f=1,2,\ldots,F}$, the overlay routing decision at time $t$ is solely determined by the network state, i.e., $\bm{U}(t) = h(\bm{Q}(t))$ for some function $h(\cdot)$.

The assumption of bounded queues is not overly restrictive since the links can simply drop a packet if their buffer overflows\footnote{The steady state packet loss probabilities can be made arbitrarily small by choosing the bound $Q_{\max}$ on the buffer size to be sufficiently large. Moreover, such an assumption automatically guarantees stability of queues, and lets us focus on our key objective of controlling network delays.}.

We note that we can also consider the set-up in which each flow $f$ has an average end-to-end delay requirement, and the objective is to design an overlay routing policy that meets the requirements imposed by each flow $f$. We exclude this set-up to make the presentation easier, and instead focus only on the case where the delays summed up over flows is to be kept below a certain threshold. Details regarding this extension can be found in Section~\ref{sec:ext}.
 
Henceforth we will write $\bar{\|\bm{Q}\|}_\pi$ to denote the total steady state average queue lengths under the application of policy $\pi$, and $\bar{Q}^f_{\ell,\pi}$ to denote the steady state queue length of flow $f$ at link $\ell$. At times, we will  suppress the dependency on $\pi$ and use $\bar{\|\bm{Q}\|}$ instead of $\bar{\|\bm{Q}\|}_\pi$. Similarly for $\bar{\|\bm{Q}_\ell\|},\bar{Q}^f_\ell$, etc.

\section{Optimal Policy Design}
\begin{figure}
\includegraphics[scale=.28]{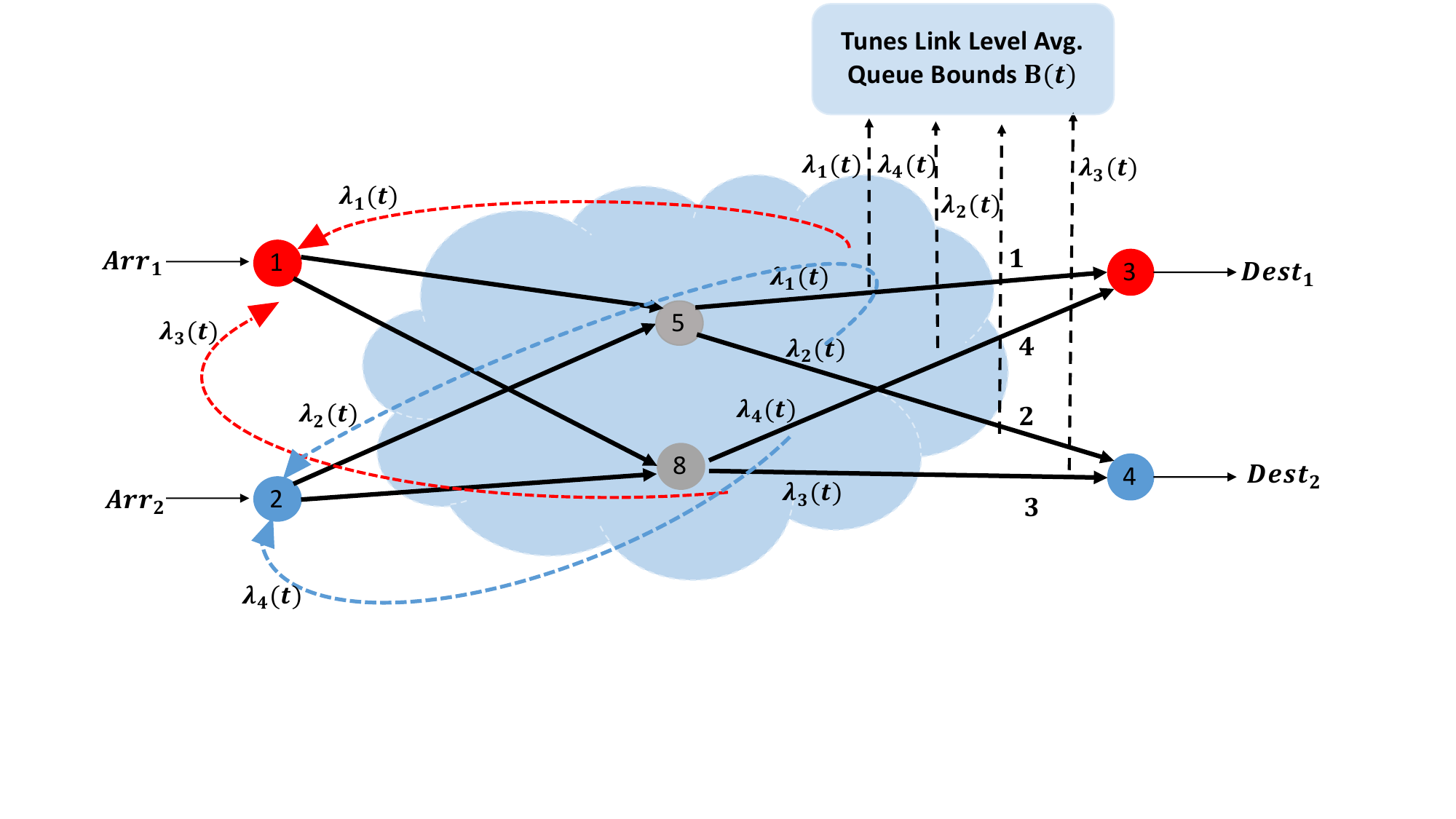}
\vspace{-.65in}
\caption{An overlay network comprising of $2$ source destination pairs $(1,3)$ and $(2,4)$. The introduction of link prices $\lambda_\ell(t)$ decouples the routing decisions at the overlay and results in a decentralized scheme. It also serves as a mediator between the overlay nodes making routing decisions, and the tuner which sets the queue thresholds $B_\ell$. }
\label{f4}
\end{figure}

We now begin our analysis when the underlay implements a static policy. We will show that it is possible to construct an optimal decentralized overlay routing policy using a $3$ layer controller as shown in Figure~\ref{f3}. The topmost layer will manipulate ``link-level average queue thresholds" $B_\ell(t)$ based on the link prices $\lambda_\ell(t)$. The link prices $\lambda_\ell(t)$ are representative of the instantaneous congestion at link $\ell$. The next two layers, link-price controller, and packet-level decision maker will collectively try to meet the bounds $B_\ell(t)$ by using a primal dual algorithm described below. The packet-level decision maker will utilize the link prices $\lambda_\ell(t)$ in order to make routing decisions $\bm{U}(t)$. It will transmit packets on routes which utilize links with lower prices. The link-price controller will then observe the congestion that results from the routing decisions $\bm{U}(t)$, and manipulate the prices accordingly in order to direct more traffic towards links on which the average queues are less than the threshold $B_\ell(t)$. The interaction between the top two layers is described by a nonlinear ordinary differential equation (ode) called the replicator equation~\cite{weibull1995evolutionary}. Fig.~\ref{f3} depicts the interactions between these layers and Fig.~\ref{f4} shows the overall structure of the scheme being used to control an overlay network. We call our 3 layer adaptive routing policy Price-based Overlay Controller (POC). We now develop these algorithms in a bottom to top fashion.
\subsection{Link-level Design}\label{sec:dba}
We notice that in order to satisfy the constraint $\sum_{\ell}\|\bar{\bm{Q}}_{\ell}\| \leq B$ imposed in the CMDP~\eqref{cmdp:del}-\eqref{cmdp:del1}, a policy $\pi$ needs to appropriately coordinate the individual link-level average queue lengths $\|\bar{\bm{Q}}_{\ell}\|$ so that their combined value is less than $B$. Such a task is difficult because the link-level average queues $\|\bar{\bm{Q}}_{\ell}\|$ have complex interdependencies between them that are described by the \emph{unknown} underlay network structure. 

In view of the above discussion, we begin by considering a simpler problem, one in which the tolerable cumulative average queue bound $B$ has already been divided into link-level ``components" $\{B_\ell\}_{\ell\in E}$ that satisfy $\sum_{\ell\in E}B_\ell = B$, and the task of overlay is to keep the average queues on each link $\ell$ bounded by the quantity $B_\ell$. An efficient scheme should be able to meet the link-level bounds $B_\ell$ in case they are achievable under \emph{some} routing scheme.

Thus, we consider the following Constrained MDP~\cite{altman}, 
\begin{align}
&\min_{\pi} \qquad 0\label{cmdp}  \\
&\mbox{s.t. }  \bar{\|\bm{Q}_\ell\|}\leq B_\ell, \quad\forall \ell \in E,\label{cmdp1}
\end{align} 
where in the above, expectation is taken with respect to the overlay policy $\pi$, randomness of packet arrivals, network link capacities and the underlay policy. Next, we derive an iterative scheme that yields a decentralized policy which solves the above CMDP, i.e., satisfies the link-level average queue length bounds in case they are achievable under some policy. The task of choosing the bounds $B_\ell$ will be deferred until Section~\ref{sec:atd}.

\subsubsection{Flow Level Policy Decomposition}\label{sec:fld}
We now show that the problem~\eqref{cmdp}-\eqref{cmdp1} of routing packets under average delay constraints at each link $\ell\in E$ admits a decentralized solution of the following form. Each underlay network link $\ell$ charges a ``holding cost" at the rate of $\lambda_\ell$ units per unit queue length $Q^f_\ell$ from each flow $f$ that utilizes link $\ell$. The holding charge collected by link $\ell$ from flow $f$ at time $t$ is thus $\lambda_\ell Q^f_\ell(t)$. The flows are provided with the link prices $\{\lambda_\ell\}_{\ell\in E}$, and they schedule their packet transmissions in order to minimize their average holding costs $\sum_{\ell}\lambda_\ell \bar{Q}^f_\ell$. Under such a scheme, the prices $\lambda_\ell$ intend to keep the network traffic from ``flooding" the links, thus enabling them to meet the average queue threshold of $B_\ell$ units. 

The routing scheme described above is \emph{decentralized}, meaning that each flow $f$ makes the routing decisions solely based on its individual queue lengths $\bm{Q}^f(t)$, and the link prices $\{\lambda_\ell\}_{\ell\in E}$, and not on the basis of global state vector $\bm{Q}(t)$. Thus, the problem of routing packets in order to meet the average queue thresholds $B_\ell$ neatly decomposes into $F$ sub-problems, one for each flow. The sub-problem for flow $f$ is to route its packets in order to minimize its individual holding cost $\sum_{\ell}\lambda_\ell \bar{Q}^f_\ell$. 

We now proceed to solve the CMDP~\eqref{cmdp}-\eqref{cmdp1}. Throughout this section we will assume that the problem~\eqref{cmdp}-\eqref{cmdp1} is feasible, i.e., there exists a routing policy $\pi$ under which the average queue lengths are less than or equal to the thresholds $B_\ell$. %Next, we consider the dual MDP corresponding to the CMDP~\eqref{cmdp:del}, which leads us to the decomposition principle discussed above.     
Let $\lambda_{\ell}\geq 0$ denote the Lagrange multiplier associated with the constraint $ \bar{\|\bm{Q}_\ell\|}\leq B_\ell$, and let $\bm{\lambda}=\left(\lambda_1,\lambda_2,\ldots,\lambda_L\right)$. If $\pi$ is a stationary randomized policy (\cite{puterman,altman}), then the Lagrangian for the problem~\eqref{cmdp}-\eqref{cmdp1} is given by~\cite{altman},
\begin{align}\label{lagrange}
\mathcal{L}(\pi,\bm{\lambda})&=\sum_{\ell}\lambda_{\ell}\left( \bar{\|\bm{Q}_\ell\|}_{\pi}- B_\ell \right)\notag\\
&=\sum_{\ell} \lambda_{\ell} \left(\sum_f \bar{Q}^f_{\ell,\pi} \right) -\sum_{\ell}\lambda_{\ell}B_\ell\notag\\
&=\sum_f \left(\sum_{\ell\in R_f} \lambda_{\ell} \bar{Q}^f_{\ell,\pi}\right)-\sum_{\ell}\lambda_{\ell}B_\ell,
\end{align} 
where the second equality follows from the relation $\|\bm{Q}_\ell(t)\|=\sum_f Q^f_\ell(t)$. The Lagrangian thus decomposes into the sum of flow $f$ ``holding costs" $\sum_{\ell\in R_f} \lambda_\ell \bar{Q}^f_\ell$, where $R_f$ is the set of links utilized for routing packets of flow $f$.

We now make the following important observation. Since the links share the service amongst their queues in a manner irrespective of the queue length $\bm{Q}_\ell (t)$, the service received by flow $f$ on link $\ell$ at any time $t$, is determined only by the capacity of link $\ell$ at time $t$, and queue length $Q^f_\ell(t)$.
Thus, the evolution of queue length process $\bm{Q}^f(t),t\geq 0$ is completely described by $\bm{U}^f(t),t\geq 0$, i.e., the routing actions chosen for flow $f$ at source node $s_f$, and the probability laws governing capacities of links $\ell$ that are used to route flow $f$ packets. Let us denote by $\pi_f$ a policy for flow $f$ that maps $\bm{Q}^f(t)$ to $\bm{U}^f(t)$.
 Since for a fixed value of $\bm{\lambda}$, the cost $\sum_{\ell\in R_f} \lambda_{\ell} \bar{Q}^f_\ell$ incurred by each flow $f$ can be optimized independently of the policies chosen for other flows, it then follows from~\eqref{lagrange} that,
\begin{lemma}\label{lemma1}
The dual function corresponding to the CMDP~\eqref{cmdp} is given by,
\begin{align}
D(\bm{\lambda}) &= \min_{\pi}\mathcal{L}(\pi,\bm{\lambda})\notag\\
&=\sum_f \min_{\pi_f} \sum_{\ell\in R_f}\lambda_{\ell}\bar{Q}^f_\ell -\sum_{\ell}\lambda_{\ell}B_\ell. \label{dualfun}, 
\end{align} 
and hence the policy $\pi$ which minimizes the Lagrangian for a fixed value of $\bm{\lambda}$ is \emph{decentralized}. % xxx  \textbf{IC 5}
\end{lemma}
We now provide a precise definition of a decentralized policy.
\subsubsection{Decentralized Policy}\label{subsubsec:decpol}
A policy $\pi$ is said to be decentralized if $\bm{U}^f(t)$, i.e., the overlay decisions regarding packets of flow $f$, are made based solely on the knowledge of flow $f$ queues $\bm{Q}^f(t)$, i.e., $\bm{U}^f(t)=h_f(\bm{Q}^f(t))$, for some function $h_f(\cdot)$, and not on the basis of global queue lengths $\bm{Q}(t)$. If $\pi$ is decentralized, we let $\pi=\otimes_{f}\pi_f$, i.e., $\pi$ is uniquely identified by describing the policies $\pi_f$ for each individual flow $f$. Upon associating a policy with the steady state measure that it induces on the joint state-action pairs, we see that the decentralized policy corresponds to the product measure $\otimes_{f}\pi_f$ on the space $\otimes_{f} \left(\mathcal{Q}_f,\mathcal{U}_f\right)$, where $\mathcal{Q}_f$ is the space on which $\bm{Q}^f(t)$ resides, and $\mathcal{U}_f$ is the action space corresponding to the routing decisions for flow $f$. We note that for a general network control problem, the optimal routing decision made by flow $f$ at time $t$ is a function of the global queue-length vector $\bm{Q}(t)=\left(\bm{Q}^1(t),\bm{Q}^2(t),\ldots,\bm{Q}^F(t)\right)$, and hence for a dynamic network control problem, a decentralized policy may not be optimal. As an example, consider the set-up of Fig.~\ref{f5} in which two flows $f_1,f_2$ inject traffic into a single stochastic link that connects nodes $3$ and $4$, and the objective is to keep the average queue lengths at link $(3,4)$ bounded by $B$. Clearly, a higher amount of traffic being injected from $f_1$ should imply that $f_2$ injects less traffic. Hence, $f_2$ must need to know the amount of traffic being sent by $f_1$ into the link $(3,4)$, or equivalently the queue length $Q^1_{(3,4)}(t)$. Similarly for $f_1$.

However, in this paper we derive a decentralized policy, and show that is optimal for the problem of dynamically routing packets where the goal is to meet a desired delay bound of $B$ units. 

Let $\bm{\lambda}^{\star}$ be the value of the Lagrange multiplier that solves the dual problem, i.e., 
\begin{align}\label{dualprob}
\bm{\lambda}^\star \in \arg\max_{\bm{\lambda} \geq \bm{0}} D(\bm{\lambda}),
\end{align}   
where $\bm{0}$ is the $|E|$ dimensional vector all of whose entries are $0$.

The average costs incurred under a policy $\pi$ can be considered as a dot product between the steady state probabilities induced over the joint state-action pair under $\pi$, and the one-step cost under state-action pairs. Hence, a CMDP can equivalently be posed as a linear program~\cite{borkar88}, and is convex. If strong duality~\cite{bertsekas} holds, then the duality gap corresponding to the problem~\eqref{cmdp}-\eqref{cmdp1}, and its dual~\eqref{dualprob} is zero~\cite{bertsekas}. A sufficient condition for strong duality is Slater's condition~\cite{bertsekas}. It is easily verified that for the problem~\eqref{cmdp}-\eqref{cmdp1}, Slater's condition reduces to the following condition: there exists a policy $\pi$ under which the link-level delays for each link $\ell$ are strictly less than the bound $B_\ell$, ie., $\bar{\|\bm{Q}_\ell\|}_{\pi}< B_\ell, \quad\forall \ell \in E$. We will assume that there is a policy $\pi$ that is strictly feasible.
 
Let us denote by $\pi^\star_f(\bm{\lambda})$ the policy for flow $f$ which minimizes the holding cost $\sum_{\ell\in R_f}\lambda_{\ell}\bar{Q}^f_\ell$. 

It then follows from the above discussion that,
\begin{theorem}\label{th:1}
Consider the CMDP~\eqref{cmdp}-\eqref{cmdp1}, and assume that there exists a policy $\pi$ under which the delay bounds are strictly satisfied, i.e., $\bar{\|\bm{Q}_\ell\|}_{\pi}< B_\ell, \quad\forall \ell \in E$. Then, there exists $\bm{\lambda}^\star=\{\lambda^\star_{\ell}\}\geq \bm{0}$ such that the CMDP~\eqref{cmdp}-\eqref{cmdp1} is solved by the policy $\pi^\star$ which implements for each flow $f$ the corresponding policy $\pi_f^\star(\bm{\lambda}^\star)$. Since the policy $\pi_f^\star(\bm{\lambda}^\star)$ requires only the knowledge of queue lengths $\bm{Q}^f(t)$ corresponding to flow $f$ in order to implement the optimal routing decision at time $t$, and not the global queue lengths $\bm{Q}(t)$, the policy $\pi^\star= \otimes_{f}\pi_f^\star(\bm{\lambda}^\star)$ is decentralized. Hence, the CMDP~\eqref{cmdp} has a decentralized solution.
\end{theorem}
The consideration of the dual problem corresponding to CMDP~\eqref{cmdp} greatly reduces the problem complexity. We were able to show that the optimal policy is decentralized. This has several simplifications discussed later in Section~\ref{sec:con}.
However, the following two issues need to be addressed in order to compute the optimal policy $\pi^\star$.
\begin{enumerate}
\item Computation of the policies $\pi^\star_f(\bm{\lambda})$ requires the knowledge of underlay topology, distribution of underlay links' capacities and statistics of packet arrival processes. These are, however, not known at the overlay. Moreover, in case the parameters are time-varying, the policy $\pi^\star_f\left(\bm{\lambda}\right)$  necessarily needs to adapt to the changes.
\item Optimal policy $\pi^\star$ in Theorem~\ref{th:1} needs to know the vector of optimal ``link prices" $\bm{\lambda}^\star$ that solve the dual problem~\eqref{dualprob}. 
\end{enumerate}
We resolve the above stated issues by employing a two timescale ``online learning" method. 
\subsubsection{Two timescale Online Learning}\label{subsubsec:tts} 
A detailed discussion of this approach can be found in Appendix A. In this section we will briefly discuss the scheme. 

For a fixed value of link prices $\bm{\lambda}$, the optimal policy for flow $f$, i.e. $\pi^\star_f(\bm{\lambda})$ can be obtained by using Dynamic Programming~\cite{bertsekasdp,puterman}. Since solving the DP optimality equations requires knowledge of network parameters, which are unknown, the optimal routing policy $\pi^{\star}(\lambda)$ can be learnt by performing Q-learning iterations~\cite{sutton}.  Q-learning algorithm keeps track of the Q-factors $V(\bm{Q}^f,\bm{U}^f),f=1,2,\ldots,F$\footnote{Since we are dealing with average cost MDPs, the algorithm we discuss is a variant of the popular Q learning algorithm that is used for solving average cost MDPs, and is called RVI Q learning. However, for brevity sake, we will denote it by Q learning.}. The Q factors represent the long-term cost associated with taking the routing action $\bm{U}^f$ when the instantaneous queue lengths of flow $f$ are equal to $\bm{Q}^f$, and thereafter following the optimal routing policy. These factors are updated using the actual realizations of the queue lengths of flow $f$. A detailed account of Q learning can be found in~\cite{sutton}.

For obtaining the optimal price vector $\bm{\lambda}^\star$, we can solve the dual problem~\eqref{dualprob} by utilizing the stochastic gradient descent method. It follows from~\eqref{lagrange} and~\eqref{dualfun} that
\begin{align*}
\frac{\partial D(\bm{\lambda})}{\partial \lambda_\ell} =  \bar{\|\bm{Q}_\ell\|}_{\pi^\star(\bm{\lambda})}-B_\ell.
\end{align*}
Under the stochastic gradient descent scheme, the average queue lengths $\bar{\|\bm{Q}_\ell\|}$ will be aprroximated by their time averages.

We will, however, combine the Q-learning and stochastic gradient iterations into a single two timescale learning algorithm. Thus, the algorithm we propose performs the following stochastic recursions on two different timescales $\alpha_t,\beta_t$, with $\beta_t=o(\alpha_t)$, i.e., $\lim_{t\to\infty}\beta_t\slash \alpha_t=0$,
\begin{align}
&V(\bm{Q}^f(t),\bm{U}^f(t)) \leftarrow \left\{V(\bm{Q}^f(t),\bm{U}^f(t))\right\} \left(1-\alpha_t\right) \notag\\
&+ \alpha_t \left\{ \sum_{\ell\in R_f}\lambda_\ell Q^f_\ell(t)+\min_{\bm{u}^f}V(\bm{Q}^f(t+1),\bm{u}^f	) - V(\bm{q}_0,\bm{u}_0)\right\}, \label{tts}\\
&\lambda_\ell (t+1) = \mathcal{M}\left\{\lambda_\ell (t) + \beta_t \left(\|\bm{Q}_\ell(t)\|-B_\ell\right)\right\}, \forall \ell \in E,\label{tts1}
\end{align}
where $\mathcal{M}(\cdot)$ is the operator that projects the price onto the compact set $[0,K]$ for some sufficiently large $K>0$, and the step-sizes satisfy $\sum_t \alpha_t=\infty,\sum_t\beta_t=\infty,\sum_t\alpha^2_t<\infty,\sum_t\beta^2_t<\infty,\lim_{t\to\infty}\frac{\beta_t}{\alpha_t}=0$. We can take $\alpha_t=1\slash t,\beta_t=1\slash t \left(1+\log t\right)$. The routing action chosen at time $t$ is $\epsilon_t$ greedy, i.e., routing action implemented for flow $f$ at time $t$ is $\arg\min_{\bm{u}^f\in \mathcal{U}^f} V(\bm{Q}^f(t),\bm{u}^f) $ with probability (w.p.) $1-\epsilon_t$ and $\bm{U}^f(t)$ is chosen uniformly at random from $\mathcal{U}^f$ w.p. $\epsilon_t$, where $\epsilon_t\to0$. See Appendix A for a detailed discussion. Henceforth, for two sequences $\alpha_t,\beta_t$ satisfying $\beta_t\slash \alpha_t\to 0$, we say $\beta_t=o(\alpha_t)$. In case the scheduler has some ``prior knowledge" about the network parameters, then the search space for the optimal action in the optimization problem $\arg\min_{\bm{u}^f\in \mathcal{U}^f} V(\bm{Q}^f(t),\bm{u}^f) $ can be reduced by restricting ourselves to only the feasible actions. As an example, if the link capacity on the ingress links is equal to $10$ packets per time-slot throughout the network, then any actions that schedule more than $10$ packets on any ingress links, can be ruled out.

The first component of the scheme is composed of Q-learning iterations~\cite{sutton}. The price iterations rely on solving the dual problem corresponding to~\eqref{dualfun} using gradient-descent iterations, in which the estimated value of the gradient $\frac{\partial D}{\partial \lambda_\ell}$ is used. The scheme can be shown to converge to optimal values, and a detailed proof is provided in Appendix A.3.
\begin{theorem}\label{th:2}[Link-Level Queue Control]
Consider the iterative scheme~\eqref{tts}-\eqref{tts1} that performs the Q-learning~\eqref{tts} and the price iterations~\eqref{tts1} simultaneously, on different timescales. Assume that for the CMDP~\eqref{cmdp}-\eqref{cmdp1}, there exists a policy $\pi$ under which the delay bounds are satisfied. Then, the iterative scheme ~\eqref{tts}-\eqref{tts1} converges, thereby yielding the optimal link prices $\bm{\lambda}^\star$, and the optimal routing policy $\pi^\star(\bm{\lambda}^\star)$. Thus, it solves the CMDP~\eqref{cmdp}-\eqref{cmdp1}, i.e., it keeps the average queue lengths on each link $\ell$ bounded by the threshold $B_\ell$.  % by ``virtualization". 
\end{theorem}
\subsection{Adaptively Tuning average Queue bounds $B_\ell$}\label{sec:atd}
The results, and the scheme~\eqref{tts}-\eqref{tts1} of Section~\ref{sec:dba} works only if the thresholds $\{B_\ell\}_{\ell\in E}$ are achievable. Thus, while developing the scheme we had implicitly assumed that the overlay knew that the link-level average queue threshold profile $\{B_\ell\}_{\ell\in E}$ could be met, and thereafter it could utilize the Algorithm of Theorem~\ref{th:2} in order to meet the delay requirements. 

But, the assumption that the overlay is able to characterize the set of achievable thresholds $\{B_\ell\}_{\ell\in E}$ is hard to justify in practice. Characterization of the set of achievable $\{B_\ell\}_{\ell\in E}$ is difficult because of the complex dependencies between various link-level delays. Furthermore, in order to calculate the average delay performance for any fixed policy $\pi$, the overlay needs to know the underlay characteristics, which we assume are unknown to the overlay. Thus, in this section, we devise an adaptive scheme that measures the link prices $\lambda_\ell$ that result when the scheme~\eqref{tts}-\eqref{tts1} is applied to the network, and utilizes them to iteratively tune the delay thresholds $B_\ell$ until an achievable vector $\bm{B}=\{B_{\ell}\}_{\ell\in E}$ satisfying $\sum_{\ell}B_\ell=B$ is obtained. The key feature of the resulting scheme \emph{is that it does not require the knowledge of the underlay network characteristics}. It increases the bounds $B_\ell$ for links with ``higher prices", and decreases $B_\ell$ for links with ``lower prices", while simultaneously ensuring that the new bounds sum up to $B$. 
 
For a fixed value of delay allocation vector $\bm{B}$, whether or not it is achievable under some policy, it can be shown that the price iterations in~\eqref{tts}-\eqref{tts1} converge. See Appendix~\ref{subsec:multiode} for a detailed proof of the same. Let us denote by $\bar{\bm{\lambda}}(\bm{B}):= \{\bar{\lambda}_{\ell}(\bm{B})\}_{\ell\in E}$ the vector comprising of link prices that result when the average queue thresholds are set at $\bm{B}$, and the iterations~\eqref{tts}-\eqref{tts1} are performed until convergence. %Thus, $\bar{\bm{\lambda}}(\bm{B})$ denotes the steady-state link prices when the scheme of Theorem~\ref{th:2} is utilized with thresholds set to $\bm{B}$.        
Next, we propose an iterative scheme to tune the delay budget vector $\bm{B}$. The scheme that we propose is based on the replicator dynamics~\cite{weibull1995evolutionary}, which has been utilized in evolutionary game theory. Let $\bm{B}(t)$ denote the vector comprising of delay budget allocations after the $t$-th iteration. In order to compute the updated value $\bm{B}(t+1)$, firstly compute the quantites $B^{a}_{\ell}(t+1)$ as follows,
\begin{align}\label{delayrecur}
& \frac{B^{a}_\ell(t+1)}{B} = \left(\frac{B_\ell(t)}{B} \vphantom{\gamma_k \left\{	\frac{B_\ell(t)}{B}  \left(\bar{\lambda}_{\ell}(B_\ell(t)  ) -\sum_{\hat{\ell}} \bar{\lambda}_{\hat{\ell}}(\bm{B}(t) )\frac{B_{\hat{\ell}} (t)}{B}  \right)	\right\} }\right.\notag\\
&\qquad \left.+ \gamma_t \left\{	\frac{B(t)}{B}  \left(\bar{\lambda}_{\ell}(\bm{B} (t)  ) -\sum_{\hat{\ell}} \bar{\lambda}_{\hat{\ell}}(B(t) )\frac{B_{\hat{\ell}} (t)}{B}  \right)	\right\}\right),\\
&\qquad \forall \ell\in E,t=1,2,\ldots.\notag
\end{align} 
Thereafter project the vector $\bm{B}^{a}(t):=\left\{B^{a}_\ell(t+1)\slash B\right\}_{\ell\in E}$ onto the $L$ dimensional simplex  
\begin{align}\notag
S=\left\{\bm{x}: \sum_{i=1}^{L}x_i=1,\quad  x_i\geq 0 \quad \forall i=1,2,\ldots,L\right\}.
\end{align}
Denoting the projection operator by $\Gamma\left(\cdot\right)$, we have
\begin{align}\label{delayrecur1}
\bm{B}(t+1)\slash B= \Gamma\left(\bm{B}^{a}(t)\slash B\right).
\end{align}
\footnote{In order to avoid introduction of unnecessary notation, we use the same index $t$ to describe evolution of $\bm{B}$ iterates and the network operation.}Such a projection is required in order to ensure that the iterates remain non-negative and satisfy the condition $\sum_{\ell}B_{\ell}(t+1)=B$.

\emph{Description of Scheme~\eqref{delayrecur}-\eqref{delayrecur1}} The quantities $\frac{B_{\ell}(t)}{B}$ denote the fraction of the cumulative delay $B$ that is allocated to link $\ell$ during iteration number $t$. While the quantity $\sum_{\hat{\ell}} \bar{\lambda}_{\hat{\ell}}(\bm{B}(t) )\frac{B_{\hat{\ell}} (t)}{B} $ is the average value of link prices $\{\bar{\lambda}_{\ell}(\bm{B}(t) )\}_{\ell\in E}$, where the prices are weighted in proportion to the fraction of delay budget allocated to the corresponding link. The proposed scheme~\eqref{delayrecur}-\eqref{delayrecur1} thus increases the fraction $\frac{B_\ell(t)}{B}$ if the price $\bar{\lambda}_{\ell}(\bm{B}(t) )$ is more than the average price, and decreases it otherwise. Hence, the price $\bar{\lambda}_{\ell}(\bm{B}(t) )$ is representative of the ``amount of traffic" that link $\ell$ can support.

We will show that the iterations~\eqref{delayrecur}-\eqref{delayrecur1} converge to an achievable vector $\bm{B}$ under which the total delay threshold $\sum_{\ell} B_\ell$ is less than $B$. %We let $\bm{\bar{\lambda}}$ be the vector with entries $\bar{\lambda}_\ell$.

We will utilize the ODE method \cite{kushner,borkarode,borkarbook} in order to analyze the discrete recursions~\eqref{delayrecur}-\eqref{delayrecur1}. We will build some machinery in order to be able to utilize the ODE method. In a nutshell, the ODE method says that the discrete recursions~\eqref{delayrecur}-\eqref{delayrecur1} asymptotically track the ode,
\begin{align}\label{delayode}
 \frac{\dot{B}_\ell}{B} =\frac{B_\ell(t)}{B}  \left(\bar{\lambda}_\ell(\bm{B}(t)) -\sum_{\hat{\ell}} \bar{\lambda}_{\hat{\ell}}(\bm{B}(t))\frac{B_{\hat{\ell}}(t)}{B}\right),\forall \ell \in E.
\end{align} 
A more precise statement is as follows.
\begin{lemma}\label{lemma:oderep}
Define a continuous time process $\tilde{\bm{B}}(t),t\geq 0,t\in\mathbb{R}$ in the following manner. Define time instants $s(n),n=0,1,\ldots$ by setting $s(n)=\sum_{m=0}^{n-1}\gamma_m$. Now let 
\begin{align*}
\tilde{\bm{B}}(s(n)) = \bm{B}(n), n= 0,1,2,\ldots.
\end{align*}
Then use linear interpolation on each interval $\left[s(n),s(n+1)\right]$ to obtain values of $\tilde{\bm{B}}(t)$ for the time interval $\left[s(n),s(n+1)\right]$.
Then for any $T>0$,
\begin{align}\label{conv}
\lim_{s \to\infty} \left(\sup_{t\in \left[s,s+T\right]}   \|\tilde{\bm{B}}(t)-\bm{B}^s(t) \|^2\right)  = 0,
\end{align}
where $\bm{B}^s(t),t\geq s$ is a solution of~\eqref{delayode} on $\left[s,\infty \right)$ with $\bm{B}^s(t) = \tilde{\bm{B}}(s)$. Thus, the recursions~\eqref{delayrecur}-\eqref{delayrecur1} can be viewed as discrete analogue of the corresponding deterministic ode~\eqref{delayode}. 
\end{lemma}
\begin{proof}
A detailed proof of~\eqref{conv} can be found in~\cite{kushner} Ch:5 Theorem 2.1. %xxx TBD: EXPLAIN WHY CAN USE IT? XXX % xxx
\end{proof}
We note that the discrete-time process $\bm{B}(t),t=0,1,\ldots$ of~\eqref{delayrecur}-\eqref{delayrecur1} is deterministic. However, the recursions~\eqref{delayrecur}-\eqref{delayrecur1} do assume that the price vector $\bar{\bm{\lambda}}(\bm{B} (t))$ is available in order to carry out the update. We will remove this assumption in Section~\ref{sec:repol}.

It now follows from Lemma~\ref{lemma:oderep} that in order to study the convergence properties of the recursions~\eqref{delayrecur}-\eqref{delayrecur1}, it suffices to analyze the detrministic ode~\eqref{delayode}.    
 
 Next, we will make some assumptions regarding the replicator ode~\eqref{delayode}, and show that the trajectory of the ode has the desired convergence properties under these assumptions. Let $\left<\cdot,\cdot\right>$ denote the dot product of two vectors. 
\begin{definition}
The ode~\eqref{delayode} satisfies the monotonicity condition~\cite{smithbook} if 
\begin{align}\label{assum1}
&\left<\frac{\bm{B}}{B}-\frac{\hat{\bm{B}}}{B}, \bar{\bm{\lambda}}(\bm{B})-\bar{\bm{\lambda}}(\hat{\bm{B}})\right><0,\\
&\qquad \forall \bm{B}\neq \hat{\bm{B}} \mbox{ s.t. }\bm{B}\slash B,\hat{\bm{B}}\slash B\in S\notag.
\end{align}
\end{definition}
The proof of the lemma below is provided in Appendix B.
\begin{lemma}\label{lemma2}
If the ode~\eqref{delayode} satisfies the monotonicity condition, then for $\bm{B}(0)$ in the interior of $S$, $\bm{B}(t)$, the solution of the ode~\eqref{delayode} satisfies $\bm{B}(t)\to \bm{B}^{\star}$, which is the unique equilibrium point of~\eqref{delayode}.
\end{lemma}
Lemmas~\ref{lemma:oderep} and~\ref{lemma2} guarantee that the iterative algorithm~\eqref{delayrecur}-\eqref{delayrecur1} that starts with a suitable value of the delay budget vector $\bm{B}(0)$, and then tunes it according to the rule~\eqref{delayrecur}-\eqref{delayrecur1} converges to a feasible value of the delay budget vector $\bm{B}^\star$. Once the feasible $\bm{B}^\star$ has been obtained, the scheme~\eqref{tts} can be used in order to route packets. In summary,
\begin{theorem}\label{th:3}[Replicator Dynamics based $\bm{B}$ tuner]
For the iterations~\eqref{delayrecur}-\eqref{delayrecur1}, we have that $\bm{B}(t)\to \bm{B}^{\star}$, and hence the average queue lengths under the scheme discussed above, are bounded by $B$. 
\end{theorem}
\begin{proof}
It follows from Lemma~\ref{lemma:oderep} that the recusions~\eqref{delayrecur}-\eqref{delayrecur1} asymptotically track the ode~\eqref{delayode} in the ``$\gamma_t\to0, t \to \infty$" limit (see Ch: 5 of \cite{kushner}, or Ch: 2 of~\cite{borkarbook}). Since the trajectory $\bm{B}(t)$ of the corresponding ode~\eqref{delayode} satisfies $\bm{B}(t)\to \bm{B}^{\star}$, the proof follows.
\end{proof}
\begin{remark}
It must be noted that whether or not the delay bound $B$ is achievable depends upon the network characteristics and traffic arrivals. Hence, if the bound $B$ is too ambitious, it may not be achievable. In this case, the condition~\eqref{assum1} will clearly not hold true, and the evolutionary algorithm will not be able to allocate the delay budgets appropriately. Thus, it is required that the network operator calculate an estimate of the value of the bound $B$ that can be achieved, using knowledge of the arrival processes, underlay link capacities etc. Another way to choose an appropriate value of $B$ can be to simulate the system performance under a ``reasonably good policy", and set $B$ equal to the average delays incurred under this policy. 
\end{remark}
\begin{remark}
We note the important role that the link-prices $\lambda_\ell$ play in order to ``signal" to the tuner about the underlay network's characteristics. Thus, for example, a link $\ell$ might be strategically located, so that a large chunk of the network traffic necessarily needs to be routed through it, thus leading to a large value of average queue lengths under \emph{any} routing policy $\pi$. Alternatively, its reliability might be low, which again leads to large value of queue lengths $Q_\ell(t)$. In either case, the quantity $B^\star_\ell$ would converge to a ``reasonably large value", which in turn is enabled by high values of prices $\lambda_\ell$ during the course of iterations~\eqref{delayrecur}-\eqref{delayrecur1}.  
\end{remark}
\subsection{Tuning $\bm{B}$ using online learning}\label{sec:repol}
The scheme proposed in Theorem~\ref{th:3} needs to compute the prices $\bar{\bm{\lambda}}(\bm{B}(t))$ in order to carry out the $t$-th update. It is shown in Appendix A that under a fixed value of $\bm{B}$, the prices $\lambda_{\ell}(t)$ converge. Thus, one way to compute the $\bar{\bm{\lambda}}(\bm{B}(t))$ is to apply Algorithm~\eqref{tts} with thresholds set to $\bm{B}(t)$, and wait for the link prices to converge to $\bar{\bm{\lambda}}(\bm{B}(t))$. 

In order to speed up this naive scheme, we would like to carry out the $B_\ell$ updates in parallel with the iterations~\eqref{tts}-\eqref{tts1}, an idea that is similar to the two timescale stochastic approximation scheme~\eqref{tts}-\eqref{tts1}. Thus, we propose the following scheme that comprises of three iterative update processes evolving simultaneously,  
\begin{align}
&V(\bm{Q}^f(t),\bm{U}^f(t)) \leftarrow \left\{V(\bm{Q}^f(t),\bm{U}^f(t))\right\} \left(1-\alpha_t\right) \notag\\
&+ \alpha_t \left\{ \sum_{\ell\in R_f}\lambda_\ell Q^f_\ell(t)+\min_{\bm{u}^f}V(\bm{Q}^f(t+1),\bm{u}^f	)-V(\bm{q}_0,\bm{u}_0)\right\},\label{eq:3ts1} \\
\notag\\
%&\bar{Q}_\ell(t+1) = \bar{Q}_\ell(t)(1-\alpha_t) + \alpha_t Q_\ell(t),\label{eq:3ts1_1}\\
&\lambda_\ell (t+1) = \mathcal{M}\left\{\lambda_\ell (t) + \beta_t \left(\|\bm{Q}_\ell(t)\|-B_\ell(t)\right)\right\}, \forall \ell \in E,\label{eq:3ts2}\\
\notag\\
& \frac{B^{a}_\ell(t+1)}{B} = \left(\frac{B_\ell(t)}{B} \vphantom{\gamma_t \left\{	\frac{B_\ell(t)}{B}  \left(\bar{\lambda}_{\ell}(B_\ell(t)  ) -\sum_{\hat{\ell}} \bar{\lambda}_{\hat{\ell}}(\bm{B}(t) )\frac{B_{\hat{\ell}} (t)}{B}  \right)	\right\} }\right.\notag\\
&\qquad \left.+ \gamma_t \left\{	\frac{B(t)}{B}  \left(\bar{\lambda}_{\ell}(\bm{B} (t)  ) -\sum_{\hat{\ell}} \bar{\lambda}_{\hat{\ell}}(B(t) )\frac{B_{\hat{\ell}} (t)}{B}  \right)	\right\}\right),\notag\\
&\qquad\qquad \forall \ell\in E,k=1,2,\ldots.\notag\\
&\bm{B}(t+1)\slash B = \Gamma\left(\bm{B}^{a}(t)\slash B\right)\label{eq:3ts3}
\end{align}
where the step sizes satisfy $\beta_t = o(\alpha_t),\gamma_t = o(\beta_t),\sum_t\alpha_t=\infty,\sum_t\beta_t=\infty,\sum_t\gamma_t=\infty,\sum_t\alpha^2_t<\infty,\sum_t\beta^2_t<\infty,\sum_t\gamma^2_t<\infty$, $\Gamma\left(\cdot\right)$ is the operator that projects the iterates onto the simplex $S$, and $\left(\bm{q}_0,\bm{u}_0\right)$ is a fixed state-action pair\footnote{As an example, we can choose $\bm{q}_0$ to be the value of the state when all the queue lengths are $0$, and the action $\bm{u}_0$ can be chosen to correspond to routing all the packets to a fixed ingress link.}. The condition $\gamma_t=o(\beta_t)$ will ensure that the $\bm{B}$ recursions view the network-wide link prices as having converged to equilibrium values for a fixed value of $\bm{B}$.  

Next, we show that such a scheme, depicted in Algorithm ~\ref{alg: end2end}, in which the $\bm{B}$ updates occur in parallel with the price and Q-learning iterations, does indeed converge. The analysis is similar to the two timescale stochastic approximation scheme discussed in Appendix A. Its proof is provided in Appendix C.
\begin{theorem}\label{th:4}[Three Time-scale POC Algorithm]
Consider the iterative scheme of~\eqref{eq:3ts1}-\eqref{eq:3ts3} that is designed to keep the average value of cumulative queue lengths less than the threshold $B$. Let the network of interest satisfy the monotonicty condition~\eqref{assum1}. Then, for the iterations~\eqref{eq:3ts1}-\eqref{eq:3ts3}, we have that almost surely, $\bm{B}(t)\to \bm{B}^{\star}$. Hence, the average queue lengths suffered under the Price-based Overlay Controller (POC) described in Algorithm~\ref{alg: end2end} are bounded by $B$. 
\end{theorem}
\subsection{Putting it all together: Three Layer POC}\label{sec:3-poc}
Algorithm~\ref{alg: end2end} is composed of the following three layers from top to bottom:
\begin{enumerate}
\item \emph{Replicator Dynamics} based $\bm{B}$ tuner~\eqref{layer1} that observes the link prices $\lambda_\ell$, and increases/decreases $B_\ell$ appropriately while keeping their sum equal to $B$.
\item \emph{Sub-gradient descent} based $\lambda_\ell$ tuner~\eqref{layer2} that is provided the delay budget $B_\ell$ by the layer $1$ described above, observes the queue lengths $\bm{Q}_\ell(t)$ and updates the prices based on the mismatch between $B_\ell$ and $\|\bm{Q}_\ell(t)\|$.
\item \emph{Q-learning} based routing policy learner~\eqref{layer3} which is provided the link prices $\lambda_\ell$, and learns to minimize the holding cost $\lambda_\ell \bar{Q}^f_\ell$. Its actions reflect the congestions on various links through the outcomes $\bar{Q}^f_\ell$, which in turn are used by layer $2$ above. 
\end{enumerate}
The above three layers are thus intimately connected and co-ordinate amongst themselves in order to attain the goal of keeping the average queue lengths bounded by $B$. This is shown in Figure~\ref{f3}.  
\begin{algorithm}[h]
\caption{Price-based Overlay Controller (POC) }
\begin{algorithmic}
\label{alg: end2end}
\STATE Fix sequences $\alpha_t,\beta_t,\gamma_t$ satisfying conditions of Theorem~\ref{th:4}. Initialize $\bm{\lambda}(0)>\bm{0}$\footnotemark, $\bm{B}(0)\slash B\in S$, where $S$ is the $L$ dimensional simplex. Perform the following iterations.
\STATE 1.) Each source node $s_f$ updates its Q-values using,
\begin{align}\label{layer3}
&V(\bm{Q}^f(t),\bm{U}^f(t)) \leftarrow \left\{V(\bm{Q}^f(t),\bm{U}^f(t))\right\} \left(1-\alpha_t\right) \notag\\
&+ \alpha_t \left\{ \sum_{\ell}\lambda_\ell(t) Q^f_\ell(t)\vphantom{\min_{\bm{u}^f}V(\bm{Q}^f(t+1),\bm{u}^f	)-V(\bm{q}_0,\bm{u}_0)}\min_{\bm{u}^f}V(\bm{Q}^f(t+1),\bm{u}^f	)-V(\bm{q}_0,\bm{u}_0)\right\}, %\tag*{Q-learning for $\pi^\star_f(\lambda)$}t
\end{align}
and the routing scheme for each flow $f$ is $\epsilon_t$ greedy.
\STATE 2.) Overlay updates the link prices according to,
\begin{align}\label{layer2}
\lambda_\ell (t+1) = \mathcal{M}\left\{\lambda_\ell (t) + \beta_t \left(\|\bm{Q}_\ell(t)\|-B_\ell(t)\right)\right\}, \forall \ell \in E.
\end{align}
\STATE 3.) Overlay adapts the link-level delay requirements for each link $\ell\in E$ according to,  
\begin{align}\label{layer1}
& B^{a}_\ell(t+1)\slash B = B_\ell(t)\slash B \vphantom{\gamma_t \left\{	\frac{B_\ell(t)}{B}  \left(\bar{\lambda}_{\ell}(B_\ell(t)  ) -\sum_{\hat{\ell}} \bar{\lambda}_{\hat{\ell}}(\bm{B}(t) )\frac{B_{\hat{\ell}} (t)}{B}  \right)	\right\} }\notag\\
& + \gamma_t \left\{	\frac{B(t)}{B}  \left(\bar{\lambda}_{\ell}(\bm{B} (t)  ) -\sum_{\hat{\ell}} \bar{\lambda}_{\hat{\ell}}(B(t) )\frac{B_{\hat{\ell}} (t)}{B}\right)	\right\},\notag\\
&\bm{B}(t+1)\slash B = \Gamma\left(\bm{B}^{a}(t)\slash B\right).
\end{align}
\end{algorithmic}
\end{algorithm}
\footnotetext{$\bm{0}$ is the vector with entries equal to $0$, and the inequality is to be taken componentwise.}
\section{Complexity of the POC and Possible Extensions}\label{sec:ext}
\subsection{ Convergence rate and complexity of the POC}
Convergence rates of stochastic approximation algorithms are well studied by now, and a detailed discussed can be obtained in Ch:10 of~\cite{kushner2012stochastic}, or Ch:8 of~\cite{borkarbook}. For a stochastic approximation algorithm that employs step-sizes $\alpha_t$, the normalized error, i.e. the difference between the parameter value at time $t$ and the convergent value normalized by $\sqrt{\alpha_t}$, is normally distributed with mean $0$. However, non-asymptotic analysis of stochastic approximation algorithms employed for reinforcement learning problems is still an ongoing research~\cite{dalal2018finite}. In reinforcement learning context, if the agent utilizes a ``cleverly designed scheme", e.g. upper confidence bounds (UCB) in order to balance exploratory-exploiratory trade-off, then the regret is proportional to $\sqrt{AS}$, where $A,S$ denote the cardinalities of action and state spaces. Thus, in our set-up, unless one resorts to function approximation using neural networks, the regret resulting from employing sub-optimal policy grows exponentially with the number of links.
\subsection{Tunnel based POC-T}
The proposed Algorithm~\ref{alg: end2end} assumes that the overlay source nodes $s_f$ have the knowledge of underlay queue lengths $\bm{Q}^f(t)$ and the set of underlay links that route its packets. To route packets efficiently when these are unknown, we propose to utilize a variant of Algorithm~\ref{alg: end2end} on the overlay network comprised of ``tunnels". 

\emph{Tunnel}: For each outgoing ingress link $\ell$ from $s_f$, and the destination node $d_f$ belonging to overlay, we say that the tunnel $\tau_{\ell,d_f}$ connects the node $s_f$ to $d_f$ in the overlay network, see Fig.~\ref{f1}. We note that the actual path taken by a packet sent on an ingress link, is composed of a sequence of underlay links that depends on the randomized routing decision taken by the underlay, and hence is not known at the overlay. 

Under the proposed scheme, each overlay source node $s_f$ maintains a ``virtual queue" $\hat{Q}^f_\tau(t)$ for each of its outgoing tunnel $\tau$. $\hat{Q}^f_\tau(t)$ are set to be equal to the ``number of packets in flight" on tunnel $\tau$, i.e., the number of packets that have been sent on tunnel $\tau$, but have not reached their destination node by time $t$.
The link prices and link-level average queue threshold requirements are now replaced by their tunnel counterparts $\lambda_\tau(t),B_\tau(t)$. Their iterations proceed in exactly the same manner as in Algorithm~\ref{alg: end2end}. The routing algorithm is much simpler, so that the routing decision made by source node $s_f$ is a function of the instantaneous tunnel queue lengths $\hat{Q}^f_\tau(t)$ and the prices $\lambda_\tau(t)$. Howevr, since the source node $s_f$ does not know instantaneous queue lengths, it cannot peform Q-learning iterations. Thus, the source nodes can resort to a heuristic routing scheme. For example, the node $s_f$ can route the packets to the tunnel $\tau$ which has the least value of $\hat{Q}^f_\tau(t)\lambda_\tau(t)$. We denote this algorithm the tunnel level POC, dubbed POC-T (tunnel level POC).
\subsection{Flow-level Average Queue Length Control}
Another important generalization is to consider the case where each flow $f$ requires its average queue lengths $\|\bar{\bm{Q}}^f\|$ to be bounded by a threshold. The POC Algorithm can be modified so that it now maintains $B^f_\ell$, i.e., link-level queue bounds for each flow $f$. Similarly, links maintain $\lambda^f_{\ell}$, i.e., separate prices for each flow $f$ and the routing algorithm for flow $f$ seeks to minimize the holding cost $\sum_{\ell\in R_f} \lambda^f_{\ell}\bar{Q}^f_\ell$. 

\subsection{Dynamic Policy at Underlay}
We can also consider the case where the underlay implements a dynamic policy such as largest queue first, or backpressure policy~\cite{tassi1}. The convergence of our proposed algorithms to the optimal policy, i.e., Theorems~\ref{th:2} and \ref{th:4} will continue to hold true in this set-up. Denote by $\pi_{ul}$ the policy being implemented at the underlay. The proof of Theorem~\ref{th:2} will then be modified by using the fact that for a fixed value of price vector $\bm{\lambda}$, the steady-state control policy applied to the network is given by $(\pi^\star(\bm{\lambda}),\pi_{ul})$. Since the resulting controlled network still evolves on a finite state-space, and the policy $(\pi^\star(\bm{\lambda}),\pi_{ul})$ is stationary, it is easily verified that the results of Section~\ref{sec:repol} hold true in this setting. 
\subsection{Hard Deadline Constraints}
 Yet another possibility is to consider the case where the data packets have hard deadline constraints, i.e., they should reach their destination within a prespecified deadline in order for them to be counted towards ``timely-throughput"~\cite{houkum13}. Real-time traffic usually makes such stringent requirements on meeting end-to-end deadlines. A metric to judge the performance of scheduling policies for such traffic is the timely-throughput metric~\cite{houkum13}. Timely throughput of a flow is the average number of packets per unit time-slot that reach their destination node within their deadline. The POC algorithm can be modified in order to maximize the cumulative timely-throughput of all the flows in the network. The packet-level router would then maximize the timely throughput minus the cost associated with using link bandwidth, which is priced at $\lambda_\ell$ units. Moreover, the topmost Replicator Dynamics based layer will not be required, and will be removed. The pricing updates will now be modified, so that they will try to meet the link capacities $C_\ell$, and not the average queue delays at link $\ell$. A detailed treatment for the case when the network is controllable can be found in~\cite{singh2016throughput}.
\subsection{Allowing Intermediate Nodes to Control Packet Routing}
In this paper we have allowed for control at only the source nodes. It may be desirable to allow some of the intermediate nodes to make routing decisions. Such an enhanced capability of the network will enable it to attain a lower value of average queue lengths. A modification to the bottommost layer will enable the POC algorithm to handle such situations. Thus, now each controllable node $i\in N$ will have to perform Q-learning updates~\eqref{eq:3ts1} in order to ``learn" its optimal routing policy. Thereafter, at each time $t$, it will make routing choices based on the instantaneous queue lengths $\bm{Q}^f(t)$, and the link prices $\bm{\lambda}$ by utilizing the optimal routing policy generated via Q-learning.
\subsection{Non i.i.d. arrivals}
We have, so far, restricted ourselves to the case of i.i.d. packet arrivals at the source nodes, though we mentioned briefly in Section~\ref{sec:pf} that we can extend our analysis to the case of ``Markovian arrivals". We now elaborate on this extension. For each flow $f$, we allow the distribution of the packet arrivals at time $t$ to depend upon its``arrival state process", denoted $S_{a,f}(t),t=1,2,\ldots$, where $S_{a,f}(t) \in \{1,2,\ldots,N_f\}$. Furthermore, if the state $S_{a,f}(t)=i$, then the arrivals have a Bernoulli distribution with parameters $N_f(i),p_f(i)$, so that the mean arrival rate when the arrival state process is in state $i$, is equal to $N_f(i)p_f(i)$. If $S_a(t):=\{S_{a,f}(t)\}_{f=1}^{F}$ denotes the ``combined" arrival state process, then, the state of each flow $f$ is now given by $\bm{X}^f(t) := \left(\bm{Q}^f(t),S_a(t)\right)$. Hence, the schedulers need the knowledge of the arrival state process $S_a(t)$ in order to make the optimal scheduling decisions.
\subsection{Node level POC}\label{subsec:node_poc}
In case the network maintains node-level queues, $\{\bm{Q}_i(t)\}_{i\in V}$ instead of separate queues for each of its outgoing links, then we can modify the POC in a straightforward manner. The POC now maintains node prices $\{\lambda_i(t)\}_{i\in V}$ and replaces the link-level delay budgets by node-level budgets $\{B_i(t)\}_{i\in V}$. The results regarding optimality of such a scheme go through. We note that since the number of links in a network can be up to $L$ times the number of nodes, such a node based scheme will yield enormous advantage with respect to faster convergence of the stochastic approximation schemes. Thus, in our simulations for the network of Fig.~\ref{simufig2}, we use node level POC.
\section{Simulations}\label{sec:simu}
We note that the size of the state-space associated with the Q-learning iterations is equal to $B_{\max}^{L}$, where $B_{\max}$ is the bound on the queueing buffer, while $L$ is the number of links in the network. Hence the POC policy suffers from the problem of state-space explosion. In order to deal with the problem of increased state-space size, we approximate the Q-function using neural networks as in~\cite{bertsekas1995neuro} when the network size is large.

\begin{figure}[h]
\includegraphics[scale=.35]{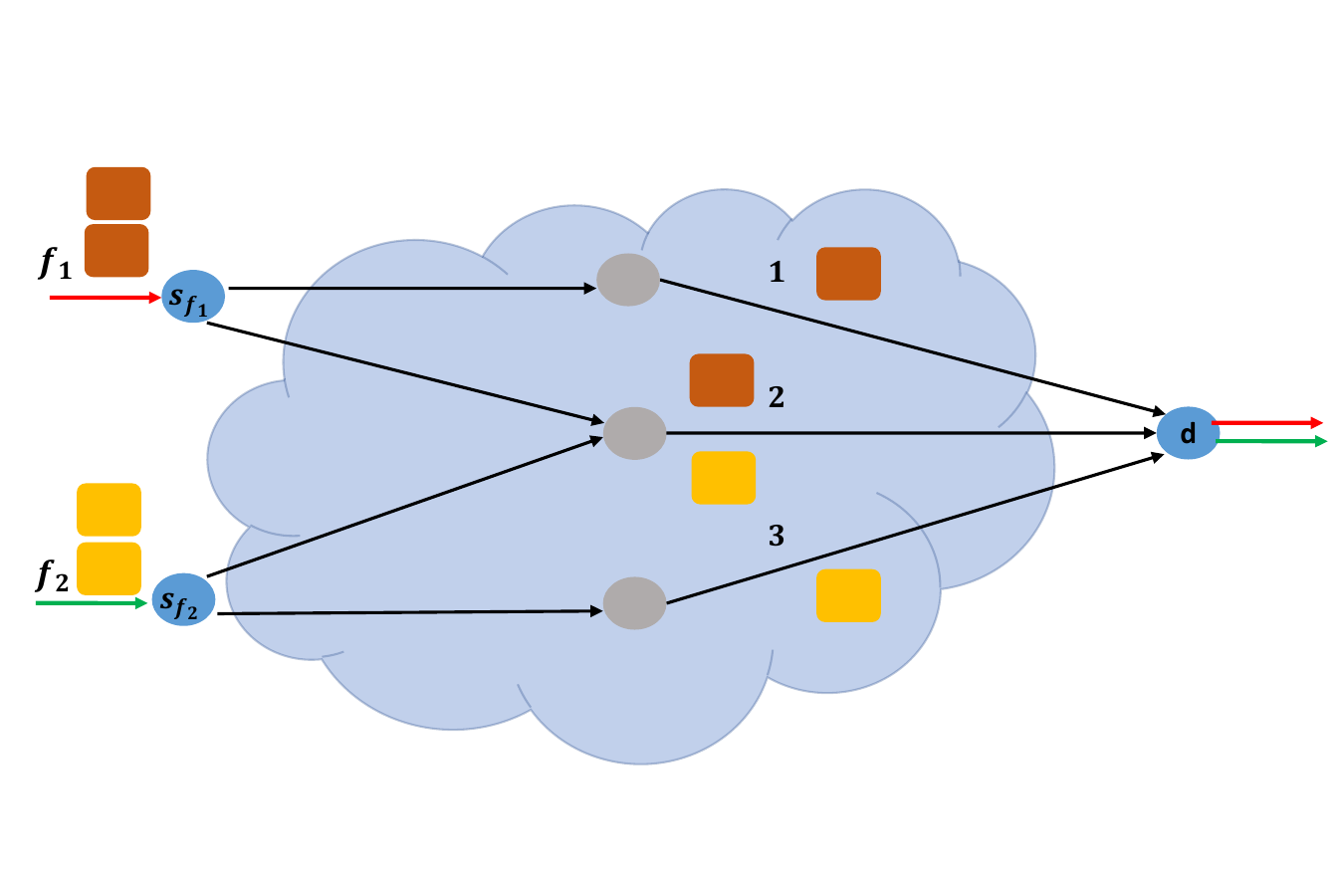}
%\vspace{-.75in}
\caption{The shaded portion of the network represents the underlay network, while the overlay is composed of two source nodes $s_{f_1},s_{f_2}$, and a single destination node $d$. The flows share the link $2$. Source node $s_{f_1}$ ($s_{f_2}$) can either route a packet on the shared link $2$, or the link $1$ ($3$) that exclusively handles traffic of $f_1$ ($f_2$). Note that $s_{f_1}$ ($s_{f_2}$) does not have access to queue lengths of $f_2$ ($f_1$). }
\label{simufig1}
\end{figure}

\begin{figure}[h]
\includegraphics[scale=.30]{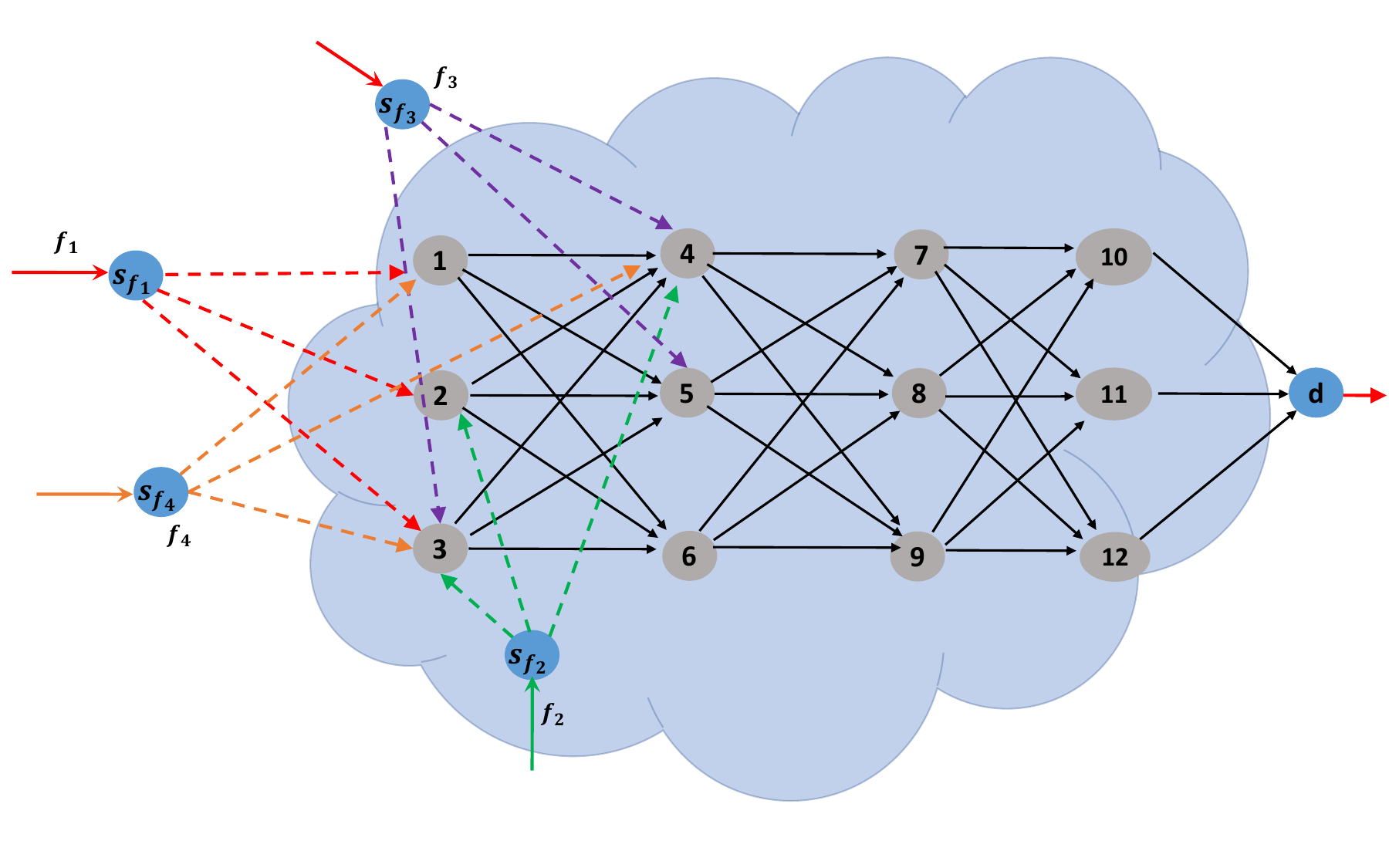}
%\vspace{-.75in}
\caption{The overlay is composed of $4$ source nodes $s_{f_1},s_{f_2},s_{f_3},s_{f_4}$, and a single destination node $d$. }
\label{simufig2}
\end{figure}

We carry out simulations for the networks shown in Fig.~\ref{simufig1} and Fig.~\ref{simufig2} in order to compare the performance of our proposed algorithms with the Overlay Back Pressure (OBP) routing algorithm that was proposed in~\cite{jones2014overlay}. We begin with a brief description of the OBP algorithm.

\begin{figure}[h]
\includegraphics[scale=.5]{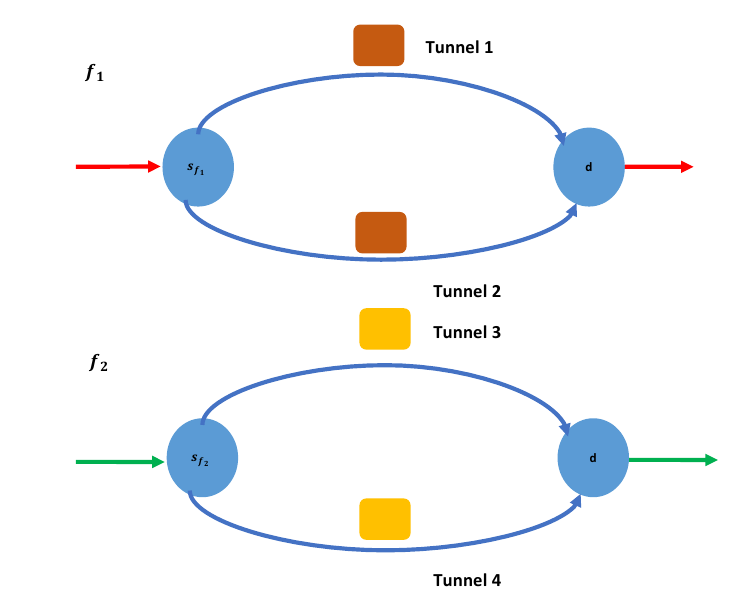}
%\vspace{-.75in}
\caption{Each source node in the network of Fig.~\ref{simufig1} has access to two tunnels on which it can send packets. Node $s_{f_1}$ can utilize tunnels $1$ and $2$, while node $s_{f_2}$ can utilize tunnels $3$ and $4$.}
\label{simufig}
\end{figure}

\emph{OBP} : For each time $t=1,2,\ldots$, define $\hat{Q}_\tau(t)$ to be the number of packets that have been sent on the tunnel $\tau$ until time $t$, but have not been received at the destination node corresponding to the tunnel $\tau$. Thus, $\hat{Q}_\tau(t)$ is the number of ``packets on flight" on tunnel $\tau$ at time $t$. At each time $t$, source node $s$ routes the packets on an outgoing tunnel $\tau$ with the least value of $\hat{Q}_\tau(t)$. The tunnels corresponding to the network of Fig.~\ref{simufig1} are shown in Fig.~\ref{simufig}.

We note that the OBP routing algorithm does not have access to the underlay queue lengths.

\subsection{Network of Fig.~\ref{simufig1}}
%For the network shown in Fig.~\ref{simufig1}, we peform simulations to judge the performance of the POC scheme summarized in Algorithm~\ref{alg: end2end}. 
We consider a slight modification with regards to the arrival and departure processes, that introduces similification to the Q-learning iterations. This is explained below.  

We assume that the packet arrivals to each source node $s_f$ follow a Poisson process, and hence the inter-arrival times between packets is exponentially distributed. The service time of packets, i.e., the time taken to successfully transfer a packet on a link is also exponentially distributed. Under the application of a policy $\pi$, the network queue lengths $\bm{Q}(t)$ evolve as a continuous-time controlled Markov process. Such a continuous-time Markov process can be converted into an equivalent discrete time Markov chain by sampling the Markov process at time-instants corresponding to arrivals and departures (real or fictitious). This technique is commonly used in the analysis of queueing systems~\cite{sennott2009stochastic}. Since for the continuous-time process, the probability of simultaneous occurance of two events is zero (e.g. an arrival and departure at the same time-epoch $t$), the continuous-time modelling assumption introduces a simplification by allowing us to consider only one state transition at each discrete time, which leads to simpler Q-learning iterations.
Let $\Lambda_f$ denote the intensity of the arrival process for flow $f$, and $R_{i}$ denote the intensity of the service process of link $i$.
Then, for the equivalent discrete-time Makov chain, there is a packet arrival for flow $f$ with probability $\Lambda_f/\left(\sum \Lambda_{\tilde{f}} + \sum R_i\right)$, while there is a departure\footnote{we note that a departure can be real or fictituous. Fictituous departures correspond to an empty queue.} (service completion) on link $i$ with a probability $R_i/\left(\sum \Lambda_{\tilde{f}} + \sum R_j\right)$. 

\begin{figure}
\includegraphics[width=8cm]{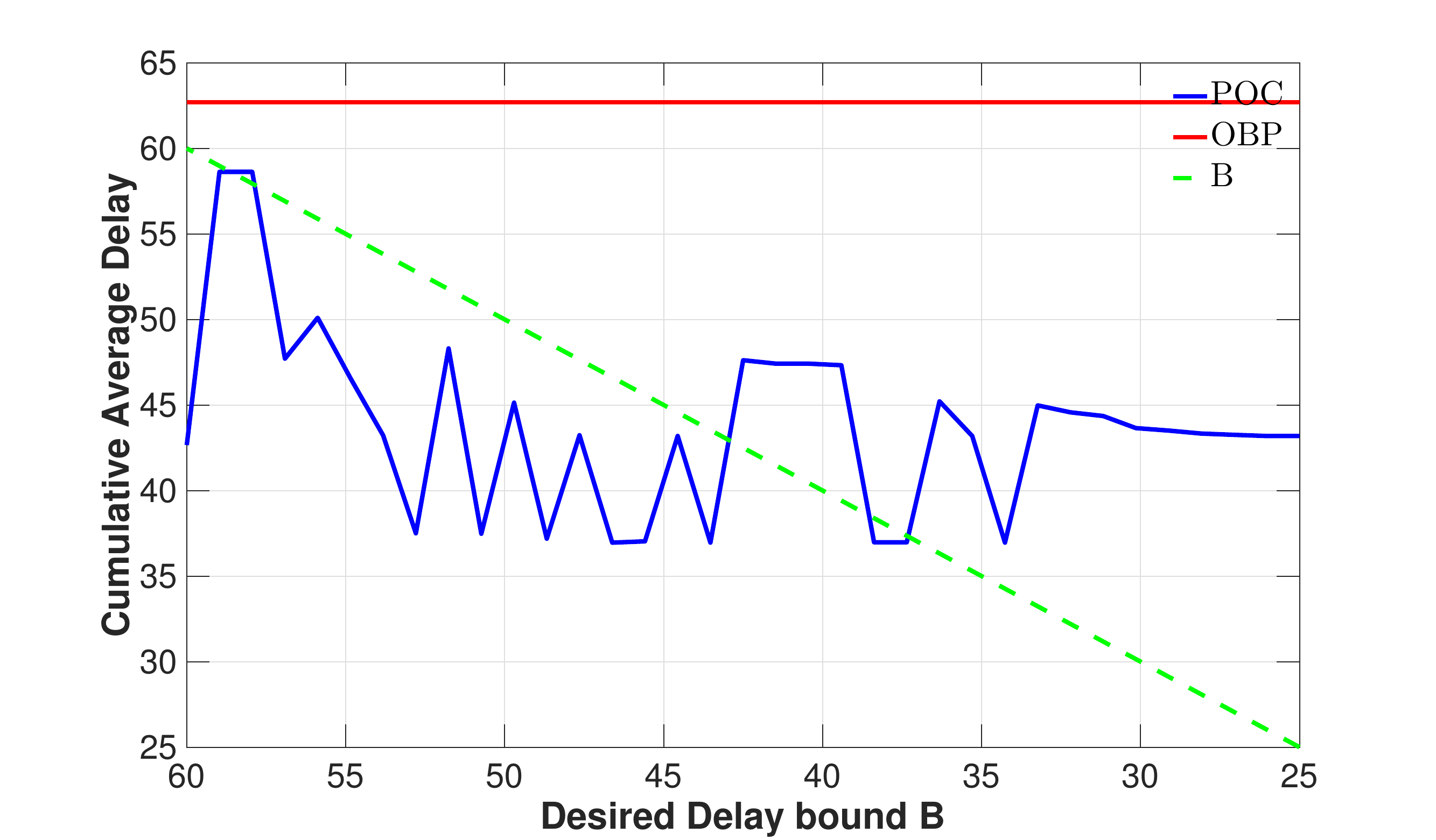}
\caption{A plot of the cumulative average delay of the POC algorithm applied to network of Fig.~\ref{simufig1} as the desired delay bound $B$ is decreased. The delay incurred by the OBP algorithm is constant. To compare the performance of POC, the desired bound $B$ is also plotted. We observe that a delay bound of $B=37$ units is achievable by the POC scheme, while delay of OBP algorithm is equal to $62.69$ units.} 
\label{s1}
\end{figure}

\begin{figure}
\includegraphics[width=8cm]{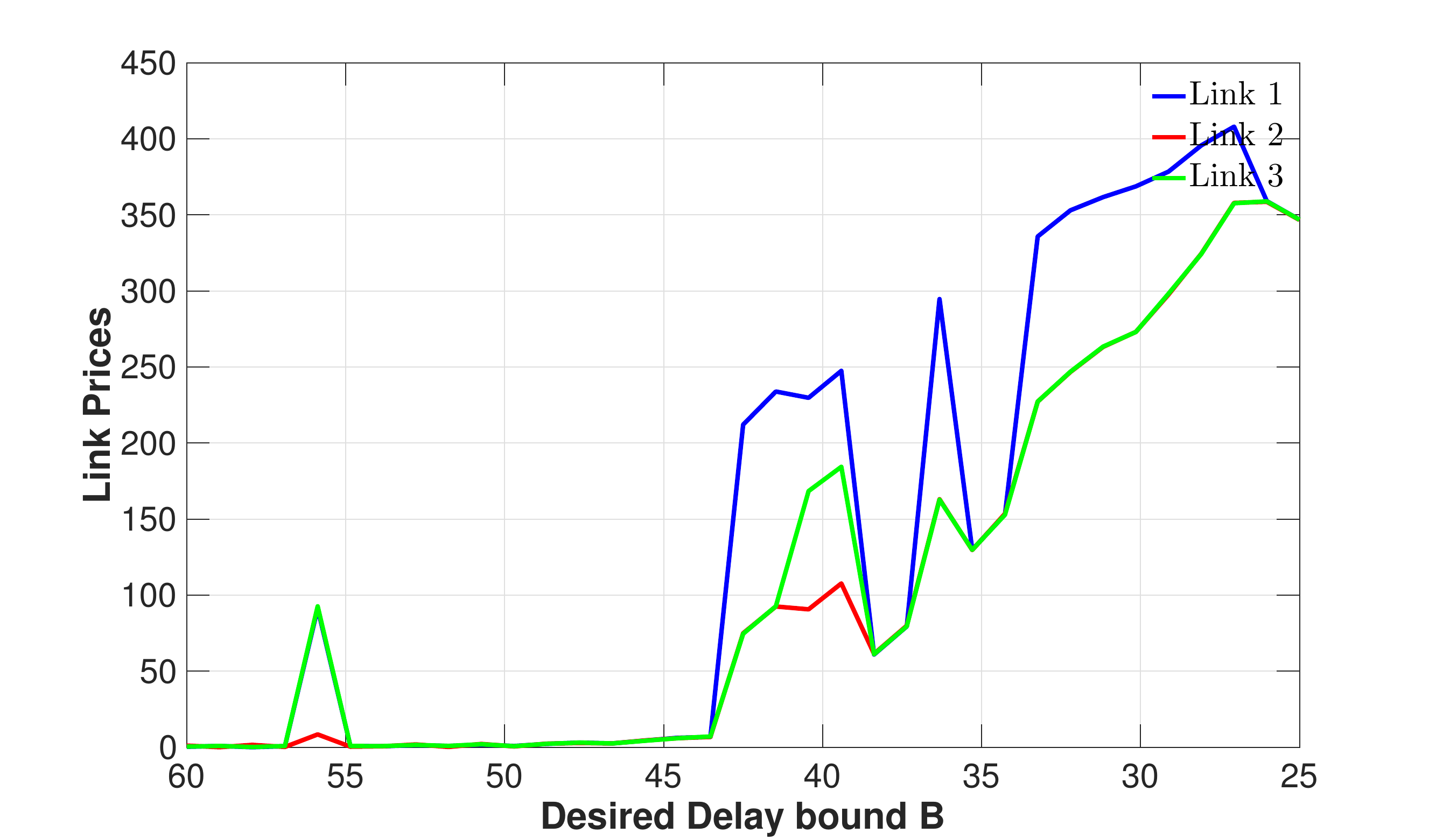}
\caption{A plot of the steady state link prices $\bm{\lambda}^{\star}(\bm{B}^{\star})$ under the POC algorithm applied to network of Fig.~\ref{simufig1} as the desired delay bound $B$ is decreased.} 
\label{s2}
\end{figure}

\begin{figure}
\includegraphics[width=8cm]{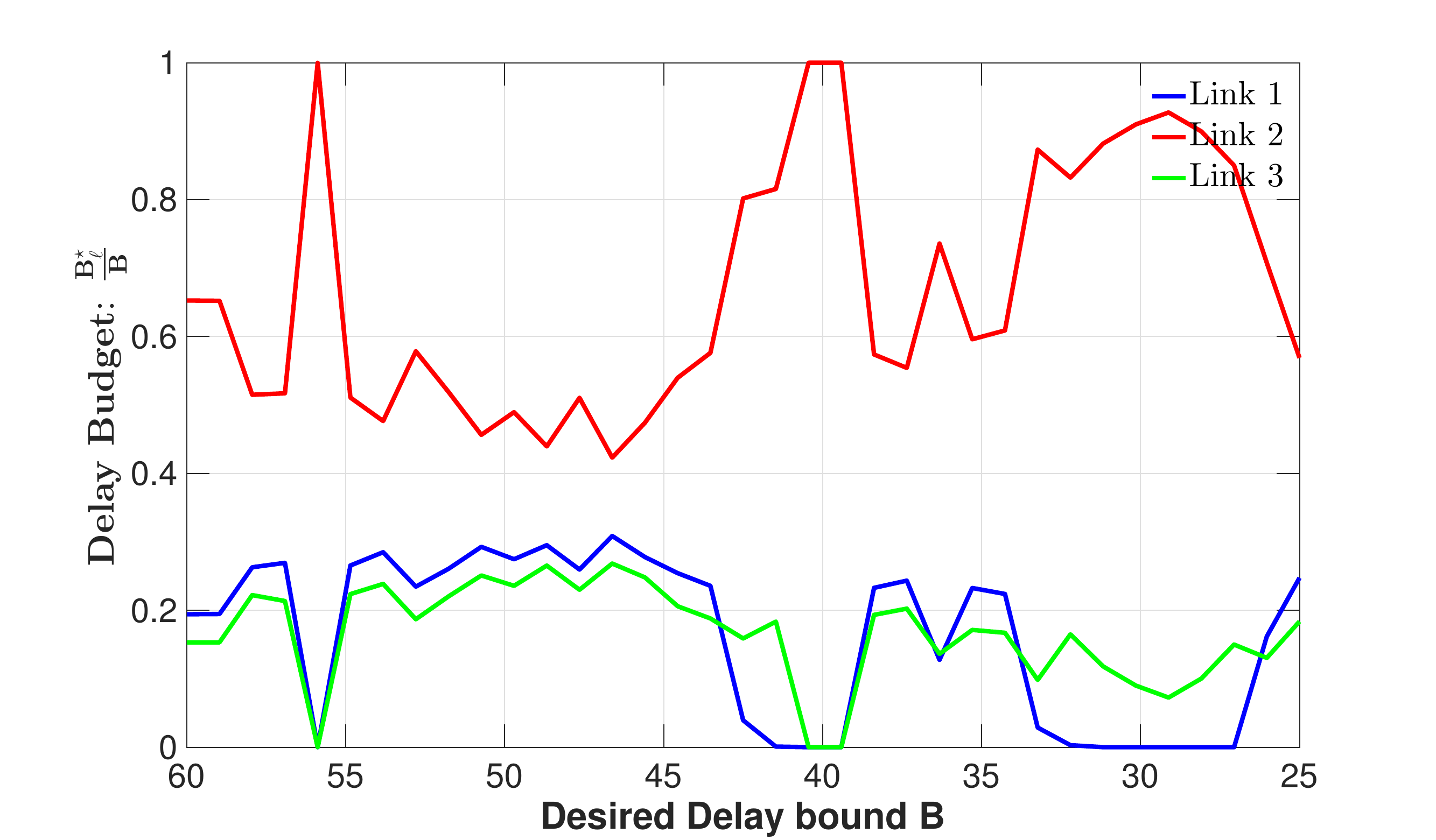}
\caption{A plot showing the steady state delay budgets $B^{\star}_\ell/B$ allocated to various links $\ell$ by the POC algorithm applied to network of Fig.~\ref{simufig1}, as the desired delay bound $B$ is decreased. We observe that since link $2$ handles traffic from both the flows, it is allocated a larger portion of the total delay bound $B$ as compared to links $1$ and $3$.} 
\label{s3}
\end{figure}

The queueing buffer capacity for each link $\ell$ is set at $B_{max}=30$ packets. We set $\Lambda_{f_1}=10,\Lambda_{f_2}=10,R_{1}=7,R_{2}=6,R_{3}=7$. The OBP algorithm incurs a delay of $62.69$ units. We then vary the delay bound parameter $B$ of the POC algorithm from $60$ units to $25$ units, and plot the cumulative average delays, steady state link prices $\bm{\lambda}^{\star}$ and steady state link budget allocations $\bm{B}^{\star}$ in Figs.~\ref{s1}-\ref{s3}. We notice that the POC algorithm significantly outperforms the OBP policy.

\begin{figure}
\includegraphics[width=8cm]{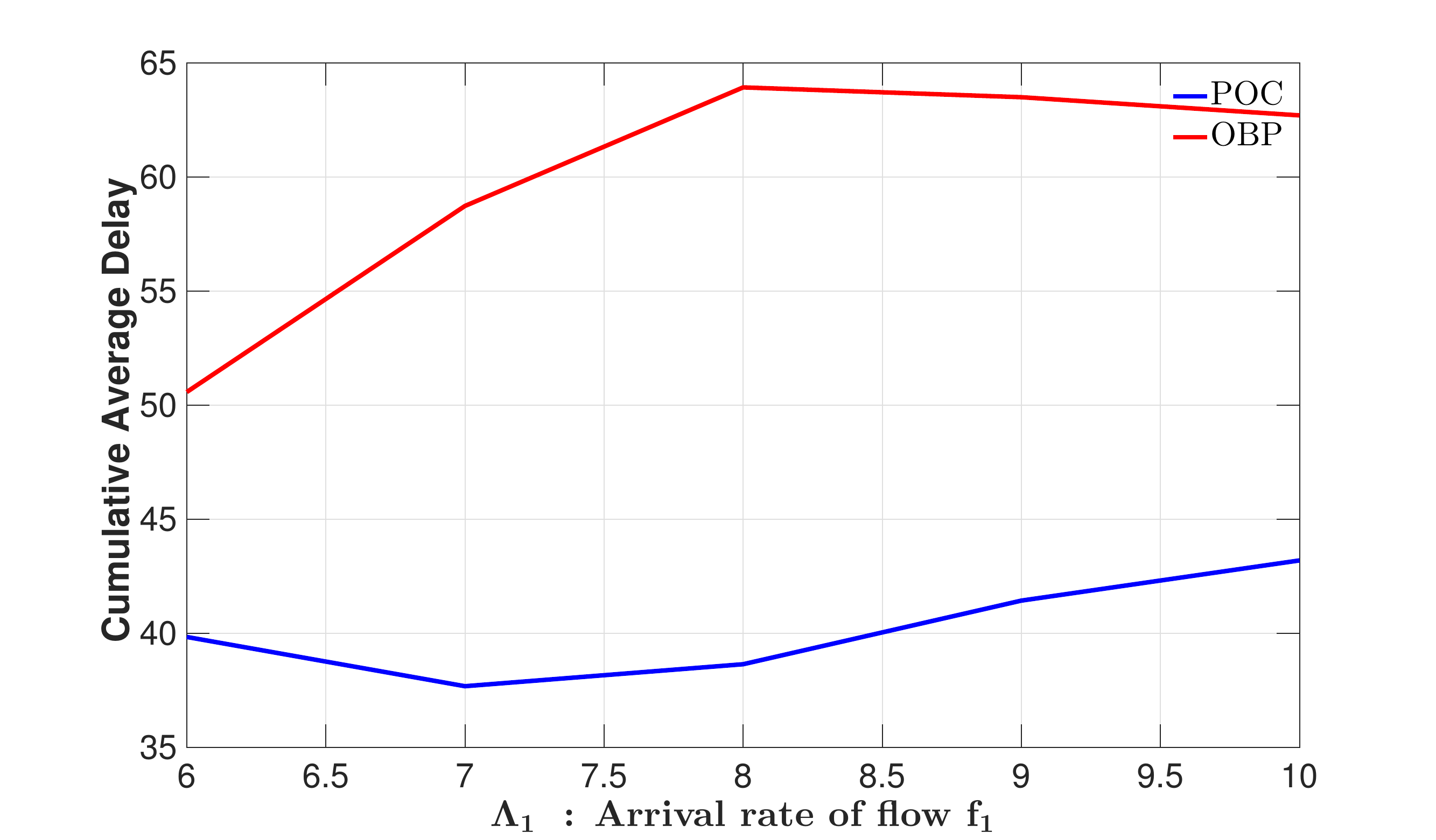}
\caption{A plot of the cumulative average delay of the POC algorithm applied to network of Fig.~\ref{simufig1} as the mean arrival rate of flow $f_1$ is increased.} 
\label{s4}
\end{figure}

\begin{figure}
\includegraphics[width=8cm]{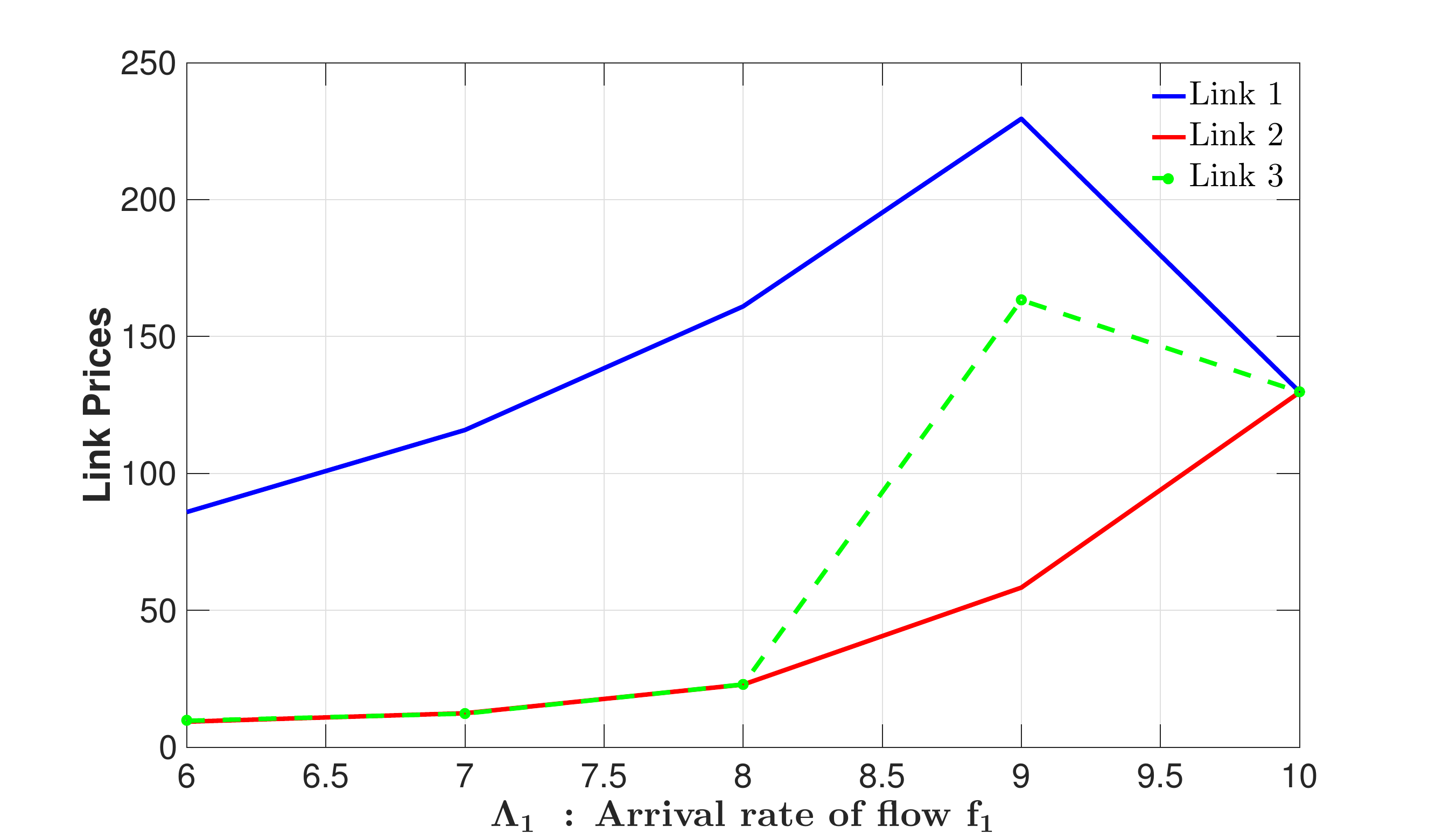}
\caption{A plot of the link prices $\lambda_\ell$ under the POC algorithm applied to network of Fig.~\ref{simufig1} as the mean arrival rate of flow $f_1$ is increased.} 
\label{s5}
\end{figure}

Thereafter, we set $\Lambda_{f_2}=10,R_{1}=7,R_{2}=6,R_{3}=7$, and vary the arrival rate $\Lambda_{f_1}$ from $6$ units to $10$ units. The results are plotted in Figs.~\ref{s4}-\ref{s5}. We observe that the POC algorithm consistently outperforms OBP policy by a significant margin.

\begin{figure}
\includegraphics[width=8cm]{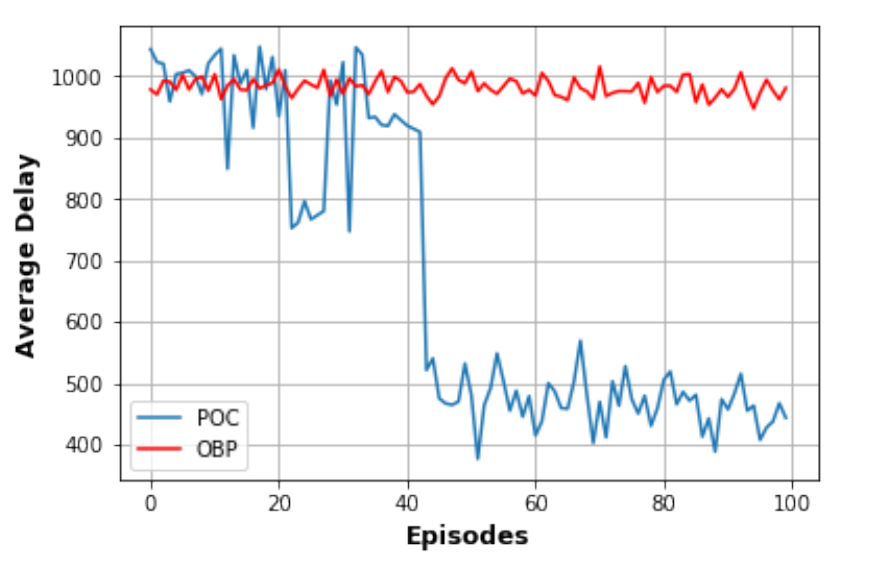}
\caption{A plot comparing the queue length processes under the OBP and the POC algorithm applied to network of Fig.~\ref{simufig2}. } 
\label{s6}
\end{figure}
\begin{figure}
\includegraphics[width=8cm]{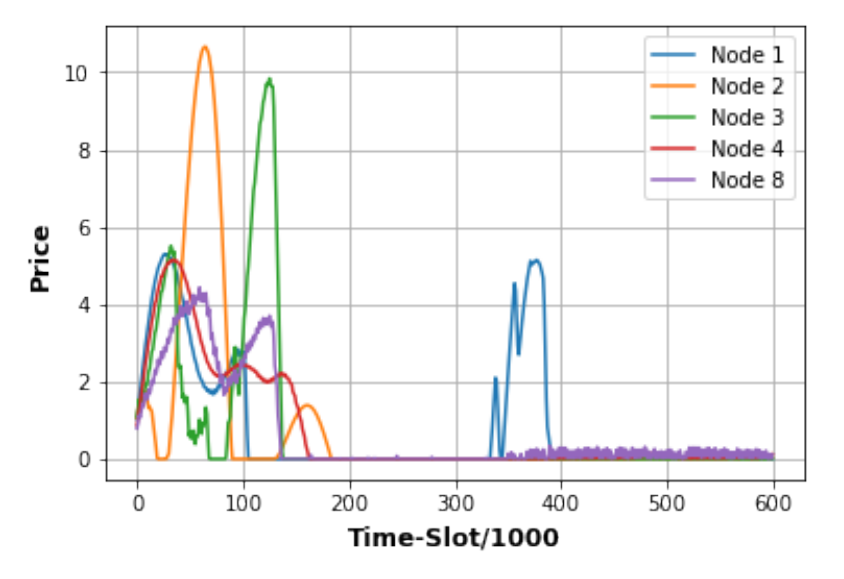}
\caption{A plot of the link prices $\lambda_\ell$ under the POC applied to network in Fig.~\ref{simufig2}. While plotting, the prices are averaged over 1000 time-slots. } 
\label{s7}
\end{figure}
\begin{figure}
\includegraphics[width=8cm]{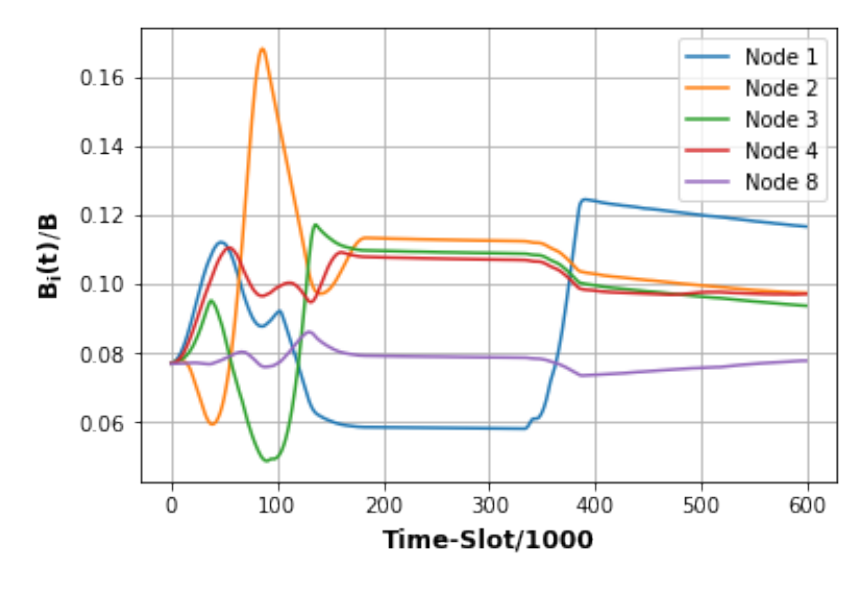}
\caption{A plot of the delay-budget $B_i(t)/B$ under the POC applied to network in Fig.~\ref{simufig2}. For the purpose of plotting, the process $B_i(t)/B$ is averaged over 1000 time-slots.} 
\label{s8}
\end{figure}
\begin{figure}
\includegraphics[width=8cm]{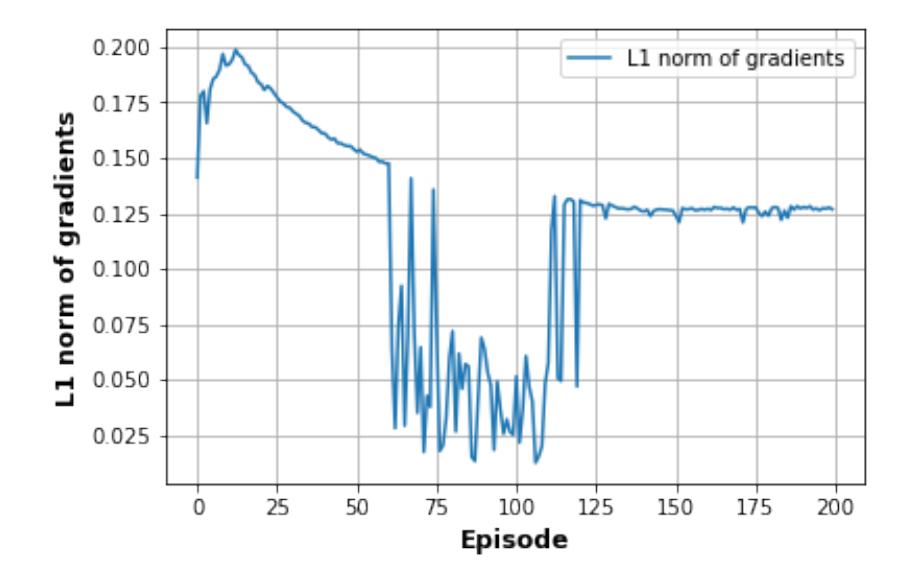}
\caption{A plot of the $L_1$ norm of the gradients, which were derived while tuning the neural network parameters, under the POC algorithm applied to network of Fig.~\ref{simufig2}.} 
\label{s9}
\end{figure}
\begin{figure}
\includegraphics[width=8cm]{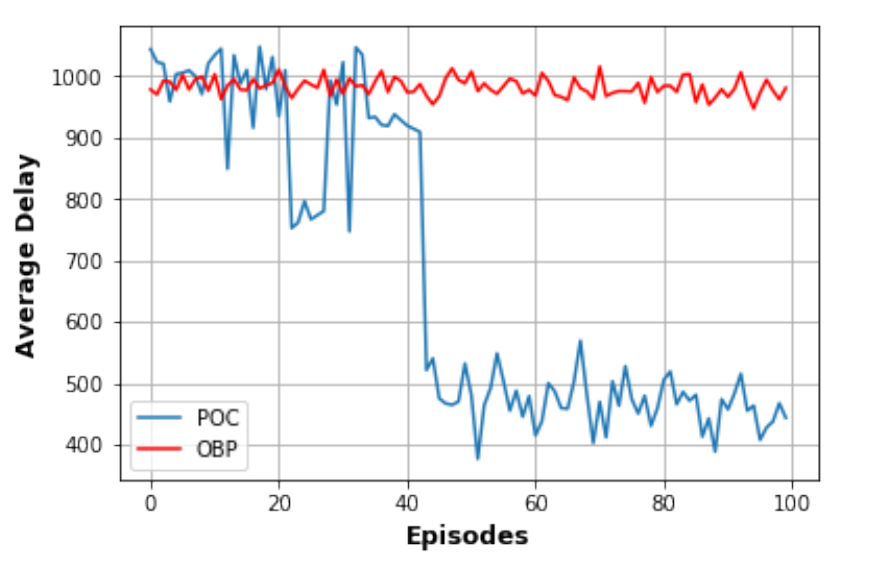}
\caption{A plot comparing the queue lengths, averaged over episode, under the OBP and the POC algorithm applied to network of Fig.~\ref{simufig2} where the packet arrivals form a Markovian pattern. We note that the ``learning phase" is of longer duration as compared to the non-Markovian arrivals case ( Fig.~\ref{s6}) because now the POC now additionally needs to learn to respond to the arrival state processes too.} 
\label{s10}
\end{figure}
\begin{figure}
\includegraphics[width=8cm]{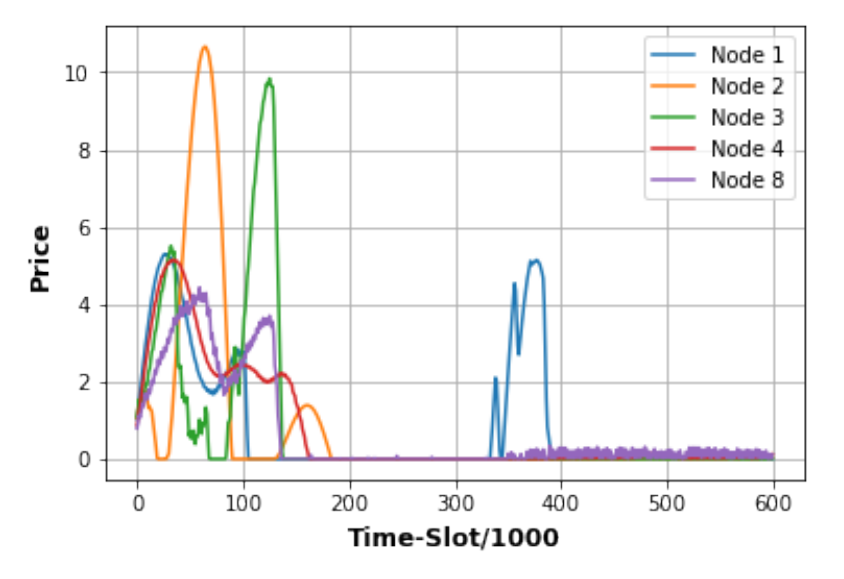}
\caption{A plot of the link prices $\lambda_\ell$ under the POC applied to network in Fig.~\ref{simufig2}, where the packet arrivals form a Markovian pattern. While plotting, the prices are averaged over 1000 time-slots. } 
\label{s11}
\end{figure}
\begin{figure}
\includegraphics[width=8cm]{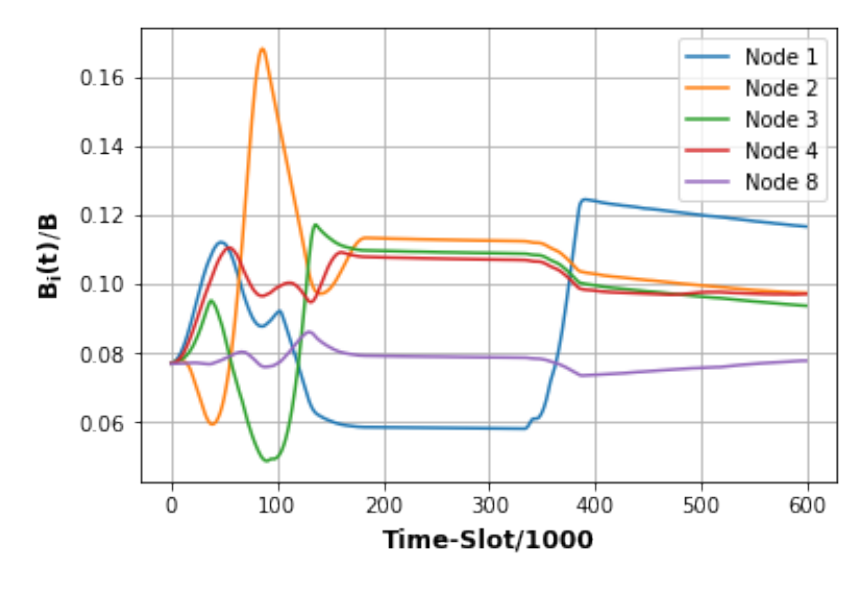}
\caption{A plot of the delay-budget $B_i(t)/B$ under the POC applied to network in Fig.~\ref{simufig2} with the packet arrivals governed by a Markov process. For the purpose of plotting, the process $B_i(t)/B$ is averaged over 1000 time-slots.} 
\label{s12}
\end{figure}
\subsection{Network of Fig.~\ref{simufig2}}
We now perform experiments on the network shown in Fig.~\ref{simufig2} that is of slightly larger scale than the one considered above. Since the network size is large, we use neural networks to approximate the Q functions for each of the flows, and we use node level POC that is discussed in Section~\ref{subsec:node_poc}. Packet arrivals are assumed to be Binomial random variables, and link states are Bernoulli. The link capacities for each link in the network was chosen uniformly at random from the set $\{1,2,3,4,5\}$. We set the episode length $\tau=3000$ time-slots. Next, we briefly describe the neural networks used for approximating the Q values for each of the 4 flows. As shown in Fig.~\ref{s6}, the OBP attained average queue lengths of around $1000$. Thus, we set the parameter $B$ for the POC, i.e. the desired threshold on average queue lengths, to $800$.

\textit{Neural Network Details} We use batch normalization as in~\cite{ioffe2015batch}, which is then followed by a linear layer. We then use a layer comprising of hard tanhyperbolic activation function with lower and upper thresholds set at $0$ and $2$ respectively, i.e. 
\begin{align}
\text{HardTanh}(x) = \begin{cases}
0 & \text{ if } x < 0 \\
2 & \text{ if } x > 2 \\
x & \text{ otherwise } .\\
\end{cases}
\end{align}
It is then followed by a layer comprising of sigmoid activations. The neural networks were implemented in PyTorch~\cite{paszke2017automatic}.

\textit{Neural Network Parameter Update Scheme} We keep the routing policy fixed for each flow $f$ within individual episodes. We set $\epsilon_t = \frac{1}{t+2}$, i.e. during episode $t$, the neural routing policy chooses the action that maximizes the Q value with a probability $1- \frac{1}{t+2}$, while it chooses from amongst the set of ingress links uniformly at random with a probability $\frac{1}{t+2}$. At the end of each episode, we randomly sample the network state values at 100 time-slots, and utilize them in order to tune the parameters of neural networks via stochastic gradient descent, with a constant step-size of $.01$. The experiment was carried out on a laptop with 2.7 GHz Intel Core i5 processor, and it took around 3 seconds to execute the computations involved in simulating the data network and those involving the neural network updates of the 4 flows. Moreover, we used constant step sizes of $10^{-5}$ and $.5\times10^{-5}$ for price updates and $B_\ell(t)$ updates respectively.

\textit{Discussion of Results} We observe from the plot in Fig.~\ref{s6}, that the POC outperforms the OBP within a few number of episodes. A plot of the prices in Fig.~\ref{s7} shows the active role played the price tuning-component in achieving promising results. Throughout the experiment, the price process is able to communicate the congestion levels to the delay budget tuner, which then adjusts the $B_\ell(t)$ accordingly. As an example,, we observe that there is a spike in the price between time-slots 1-100 for node $2$. Consequently, the delay budget allotted to node 2 also increases in order to accommodate the possibility that the channel capacity on the outgoing links for node $2$ might be ``bad".
 
A plot of the $L_1$ norm of the gradients of the neural networks in Fig.~\ref{s9} shows that the neural networks ``learn" at a decent rate. The non-monotonous nature of the $L_1$ norm around episode 75 may be explained by the ``sudden change" in the link-level prices Fig.~\ref{s7}, since the instantaneous ``reward" $\sum_{\ell\in R_f}\lambda_\ell Q^f_\ell(t)$ earned by the policy for flow $f$, depends upon the link-level prices. 

\textit{Markovian Arrivals} We then let the arrivals be governed by a Markov process that assumes binary values. The arrivals in each state have Bernoulli distribution, with the distribution parameters are different for different states.  We set episode length equal to $2000$ time-slots, the delay threshold $B$ for the POC equal to $800$. Figs.~\ref{s10}-\ref{s12} show the plots of queue-lengths, link-level prices, and delay budgets for some of the network links. the POC significantly outperforms the OBP. We observe that as expected, a spike in the prices for a node is followed by a spike in the corresponding $\frac{B_i(t)}{B}$ curve. Moreover, since the POC is able to meet the desired bound of $800$, the prices decay towards the end.
\section{Conclusions}\label{sec:con}
We developed and analyzed optimal overlay routing scheme in a system where the network operator has to satisfy a performance bound of average end-to-end delays. The problem is challenging because the overlay does not know the underlay characteristics. We proposed a simple decentralized algorithm based on a $3$ layered adaptive control design. The algorithms could easily be tuned to operate under a vast multitude of information available about the underlay congestion. In Theorem~\ref{th:1} we obtained a key flow level decomposition result which has several advantages that are listed below. 

Since an overlay routing specifies an action for each possible value of network state, the number of \emph{centralized} routing policies is given by $\prod_{f=1}^F |\mathcal{U}^f|^{Q_{\max}^{N_f}}$, where $N_f$ is the number of links which route flow $f$'s packets. On the other hand, the number of decentralized policies is equal to $\sum_{f=1}^F |\mathcal{U}^{f}|^{Q_{\max}^{N_f}}$. Thus, the search for an optimal policy in the class of decentralized policies is computationally less expensive as compared to centralized policies. Moreover, the optimal policy can be solved in a parallel and distributed fashion, in which each source node $s_f$ performs a search for flow $f$'s optimal policy using online learning techniques.

In order to implement a decentralized policy, each source node $s_f$ needs to know only its own queue lengths $\bm{Q}^f(t)$. This eliminates the need to share the network queue lengths $\bm{Q}(t)$ globally amongst all the overlay nodes, thus saving significant communication overheads.

We also propose a heuristic scheme in case the flows do not know the links that are used to route their traffic. Simulation results show that the proposed schemes significantly outperform the existing policies. 

\bibliographystyle{IEEEtran}
\bibliography{main}
\cleardoublepage
\pagenumbering{arabic}% resets `page` counter to 1
\renewcommand*{\thepage}{\arabic{page}}
\appendices
\section{Using Two Time-scale based Online Learning Algorithm for solving Link-level control problem~\eqref{cmdp}-\eqref{cmdp1}}\label{subsec:multiode}
Continuing our discussion from the end of Section~\ref{sec:dba}, we now address the problem of solving for the optimal policy $\pi^\star_f(\bm{\lambda})$, and the optimal link prices $\bm{\lambda}^\star$ \emph{without} the knowledge of network parameters. We provide online learning algorithms that separately solve these two problems of obtaining a) optimal policy $\pi^\star_f(\bm{\lambda})$ for a fixed value of link prices $\bm{\lambda}$, and b) vector of optimal link prices $\bm{\lambda}^{\star}$. Later, in Section~\ref{subsec:2ode} we combine them into a single two timescale algorithm that solves both of these problems.
\subsection{Employing Online Learning to Obtain $\pi^\star_f(\bm{\lambda})$}\label{subsec:ql}
We can utilize online learning methods such as Relative Q-learning~\cite{abounadi2001learning}, henceforth referred to just as Q-learning, in order to obtain $\pi^\star_f(\bm{\lambda})$, i.e., the policy that minimizes the holding cost $\sum_{\ell}\lambda_{\ell}\bar{Q}^f_\ell $ for flow $f$. Note that we have suppressed the dependency of $V(\cdot,\cdot,\cdot)$ on the flow $f$ in order to keep the notation simple.

In order to make the exposition simpler, we discuss the Dynamic Programming equations corresponding to the $\beta\in\left(0,1\right)$ discounted infinite horizon holding cost minimization problem. Let $\mathcal{U}$ denote the set of routing actions available to the algorithm. The Dynamic Programming~\cite{bertsekasdp} equation that needs to be solved in order to obtain the optimal routing policy that minimizes the $\beta$ discounted holding cost is,
\begin{align}\label{eq:dp:route}
V^{\star}_{\bm{\lambda}}(\bm{q},\bm{u}) &= \sum_{\ell}\lambda_{\ell}q_\ell +\beta \mathbb{E}\min_{\bm{\tilde{u}}}V^{\star}(\left(\bm{q}-\bm{D}\right)^{+}+\bm{A},\bm{\tilde{u}}),\\
&~\qquad\qquad\forall \bm{q}\in Q_{\max}^L,\bm{u}\in\mathcal{U}\notag,  
\end{align}
where the random vectors $\bm{A},\bm{D}$ are distributed according to the the number of packet arrivals and departures under the routing action $\bm{u}$, and expectation is taking with respect to randomness of $\bm{A},\bm{D}$. The quantity $V_{\bm{\lambda}}^{\star}(\bm{q},\bm{u})$ denotes the cost associated with a system starting in state $\bm{q}$, and applying the action $\bm{u}$ during the first time step, and thereafter following the optimal routing policy. The subscript $\bm{\lambda}$ is used since the cost incurred by the system depends upon the link price vector $\bm{\lambda}$. Once the equation~\eqref{eq:dp:route} has been solved for $V_{\bm{\lambda}}^{\star}:=\{V_{\bm{\lambda}}^{\star}(\bm{q},\bm{u})\}_{\bm{q}\in Q_{\max}^L,\bm{u}\in\mathcal{U}}$ the optimal policy implements the action $\bm{u}$ that minimizes the quantity $\mathbb{E}\min_{\bm{\tilde{u}}}V_{\bm{\lambda}}^{\star}(\left(\bm{q}-\bm{D}\right)^{+}+\bm{A},\bm{\tilde{u}})$, when the system state is $\bm{q}$.

The Q-learning algorithm~\cite{abounadi2001learning,sutton} utilizes stochastic approximation~\cite{robbins,kushner2012stochastic,borkarbook} in order to solve the fixed point equation~\eqref{eq:dp:route}. References~\cite{tsitsiklis1994asynchronous,jaakkola1994convergence,borkarbook,kushner} show the convergence of Q-learning algorithm for the simpler case of $\beta$-discounted Q-learning, while~\cite{abounadi2001learning} analyzes a variant of the same called RVI Q learning for the infinite horizon average cost problem. Since we will be concerned with minimizing the infinite horizon average holding cost $\sum_{\ell}Q_{\ell}$, the RVI Q-learning algorithm we discuss now is taken from~\cite{abounadi2001learning}. For the sake of brevity, we will call the RVI Q learning algorithm as simply Q learning.

Under $Q$ learning, each overlay source node $s_f$ maintains the \emph{Q-values}~\cite{abounadi2001learning,sutton} $V(\bm{q}^f,\bm{u}^f)$. These Q-values are updated using online data $\left(\bm{Q}(t),\bm{U}(t)\right),t=1,2,\ldots$. Letting $V(t,\bm{q},\bm{u})$ denote the Q-value associated with the state-action pair $(\bm{q},\bm{u})$ during iteration number $t$, the Q-learning iterations are given by
\begin{align}
&V(t+1,\bm{Q}^f(t),\bm{U}^f(t)) = \left\{V(t,\bm{Q}^f(t),\bm{U}^f(t))\right\} \left(1-\alpha_t\right) \notag\\
&+ \alpha_t \left\{ \sum_{\ell\in R_f}\lambda_\ell Q^f_\ell(t)+\min_{\bm{u}^f}V(t,\bm{Q}^f(t+1),\bm{u}^f	)\right.\notag\\
&\qquad\qquad\qquad\left.\vphantom{\sum_{\ell\in R_f}\lambda_\ell Q^f_\ell(t)+\min_{\bm{u}^f}V(t,\bm{Q}^f(t+1),\bm{u}^f	)}- V(t,\bm{q}_0,\bm{u}_0)\right\}, %\tag*{Q-learning for $\pi^\star_f(\lambda)$}
\label{eq:qlearn1}
\end{align}
where $\alpha_t=\frac{1}{t}$ is the step-size at time $t$, and $(\bm{q}_0,\bm{u}_0)$ is a fixed state-action pair for which the $V$ function is assigned the value $0$ at the end of each iteration\footnote{such a deduction of $V(\bm{q}_0,\bm{u}_0)$ from each of the Q values is needed to keep the iterates bounded.}. After dropping the time argument $t$, the iterations can be written more compactly as follows,
\begin{align}
&V(\bm{Q}^f(t),\bm{U}^f(t)) \leftarrow \left\{V(\bm{Q}(t),\bm{U}(t))\right\} \left(1-\alpha_t\right) \notag\\
&+ \alpha_t \left\{ \sum_{\ell\in R}\lambda_\ell Q^f_\ell(t)+\min_{\bm{u}}V(\bm{Q}(t+1),\bm{u}	) - V(\bm{q}_0,\bm{u}_0)\right\}, 
\label{eq:qlearn}
\end{align}
\footnote{only the Q-values corresponding to the state-action pair $\left(\bm{q},\bm{u}\right)$ that is actually realized at time $t$, i.e., $\left(\bm{Q}(t),\bm{U}(t)\right)$ gets updated to $\left\{V(\bm{Q}(t),\bm{U}(t))\right\} \left(1-\alpha_t\right)+ \alpha_t \left\{ \sum_{\ell\in R}\lambda_\ell Q_\ell(t)+\min_{\bm{u}}V(\bm{Q}(t+1),\bm{u}	) - V(\bm{q}_0,\bm{u}_0)\right\}$, while the Q-values corresponding to the remaining state-action pairs is unchanged. 
 }.
The routing action at time $t$, i.e., $\bm{U}^f(t)$ is chosen as follows. With a probability $1-\epsilon_t<1$, the algorithm chooses the action $\bm{u}^{\star}\in \arg\min_{\bm{u}\in\mathcal{U}} V(t,\bm{Q}(t),\bm{u})$ that is optimal according to the current estimate of the optimal Q values. While, with a probability $\epsilon_t>0$, the algorithm chooses $\bm{u}$ from the set of allowable actions $\bm{U}$ uniformly at random. We let $\epsilon_t\to 0$, so that asymptotically the algorithm implements the optimal policy. We call this exploration-exploitation trade-off strategy $\epsilon_t$ greedy. Choosing an action, with a small probability $\epsilon_t$, from amongst the set of actions that are sub-optimal based on the current estimate of Q values ensures that each state-action pair $\left(\bm{q},\bm{u}\right)$ gets ``visited" or ``explored" infinitely often, so that the Q learning iterations converge to the optimal values $V_{\bm{\lambda}}^{\star}$~\cite{sutton}. We also let $\bm{V}(t)=\{V(t,\bm{q},\bm{u})\}_{\bm{q}\in B^L,\bm{u}\in\mathcal{U}}$ denote the vector consisting of Q values at time $t$.

Convergence of the Q-learning scheme to the optimal policy for reinforcement learning problems is well-understood by now, and a detailed convergence proof of the algorithm~\eqref{eq:qlearn1} to the optimal policy can be found in~\cite{abounadi2001learning}. We however state this result as a lemma since it will be required in convergence analysis of algorithms in later sections.
\begin{lemma}\label{lemma:bell}
The asymptotic properties of the stochastic iterations~\eqref{eq:qlearn} can be derived by studying the following ordinary differential equation (ode),
\begin{align}\label{eq:bellmanode}
\dot{\bm{V}}(t) = T(\bm{V}(t)) - \bm{V}(t),
\end{align} 
\footnote{the equation is to be read componentwise}where the vector $\bm{V}(t)$ is composed of entries $V(t,\bm{q},\bm{u})$, and the operator $T$ is defined as follows
\begin{align*}
T(\bm{V}(t))(\bm{q},\bm{u}) &= \sum_{\ell}\lambda_{\ell}q_\ell +\ \mathbb{E}\left\{\min_{\bm{\tilde{u}}\in\mathcal{U}}V(t,\bm{q}+\bm{A}-\bm{D},\bm{\tilde{u}})\right\}\\
& - V(t,\bm{q}_0,\bm{u}_0),\forall \bm{q}\in B^L,\bm{u}\in \mathcal{U},
\end{align*}
where $\bm{A},\bm{D}$ are random vectors that have same distribution as the arrivals and departures that occur when queue length is equal to $\bm{q}$, and control action $\bm{u}$ is chosen. The expectation in the definition of $T(V)$ is taken with respect to randomness of packet arrivals and network link capacities, and $\left(\bm{q}_0,\bm{u}_0\right)$ is a fixed state-action pair for which the Q value is assigned the value $0$ at the end of each iteration.

For the ode~\eqref{eq:bellmanode}, the vector $V_{\bm{\lambda}}^{\star}$ comprising of optimal Q-values is unique globally asymptotically stable equilibrium point for the ode~\eqref{eq:bellmanode}, the ode~\eqref{eq:bellmanode}, and hence the Q-learning iterations~\eqref{eq:qlearn} converge to $V_{\bm{\lambda}}^{\star}$, thus yielding the optimal routing policy.   
\end{lemma}
\begin{proof}
See~\cite{abounadi2001learning}.  
\end{proof}
We note that in order to avoid introduction of unnecessary notations, we have used the same symbol $\bm{V}(t)$ in order to denote the Q-values for \emph{discrete time stochastic iterations}~\eqref{eq:qlearn}, and also the \emph{continuous time} solution of the \emph{deterministic} ode~\eqref{eq:bellmanode}. In order to avoid confusion, we will explicitly mention which of the above two processes is being referred to.

We note that the Q-learning algorithm as discussed above learns the optimal policy, i.e., a mapping from the instantaneous queue lengths $\bm{Q}^f(t)$ to a routing action $\bm{U}^f(t)$. It does not require the knowledge of the dynamics of queue lengths evolution in the network, and this is precisely the reason why we employ Q-learning, because the queue dynamics are unknown to the policy at the overlay.
\subsection{Obtaining $\bm{\lambda}^\star$ using Subgradient Descent Method}\label{subsec:pri}
Next, we address the issue of obtaining the optimal value of vector of link prices $\bm{\lambda}^\star$ that solve the dual problem
\begin{align}\label{eq:dualproblem}
\max_{\bm{\lambda} \geq 0} D(\bm{\lambda}).
\end{align}
Firstly we recall the definition of $\bm{\lambda}^{\star}$. Assume that there exists a policy under which the delay bounds $\bar{\|\bm{Q}_\ell\|}\leq B_\ell$ are satisfied.
Now, if the link prices are set at $\bm{\lambda}^\star$, and if each flow $f$ applies the policy $\pi_f^{\star}(\bm{\lambda}^{\star})$ that minimizes its holding cost $\sum_{\ell\in R_f}\lambda^{\star}_{\ell}\bar{Q}^f_\ell$, then the delay requirements $\bar{\|\bm{Q}_\ell\|}\leq B_\ell, \quad\forall \ell \in E$ are met.
 
Let us, for the time being, assume that for a given value of link prices $\bm{\lambda}$, the policies $\pi_f^\star(\bm{\lambda}),f=1,2,\ldots,F$ have been obtained. This could be achieved, for example, by performing the Q-learning iterations at each source node $s_f$ until convergence. Since
\begin{align}\label{eq:dualprob}
D(\bm{\lambda}) = \mathcal{L}(\pi^\star(\bm{\lambda}),\bm{\lambda}),
\end{align}
and from~\eqref{lagrange} we have,
\begin{align}\label{eq:deri}
\frac{\partial \mathcal{L} }{\partial \lambda_\ell} = \bar{\|\bm{Q}_\ell\|}_{\pi} - B_\ell,\quad \forall \ell \in E,
\end{align} 
%\footnote{note that we have suppressed the dependency of steady-state average queue lengths $\bar{\|\bm{Q}_\ell\|}$ on the policy $\pi$ that is being applied}we can use the gradient descent method~\cite{bertsekas} to compute $\lambda^\star$ via the following iterations,
\begin{align}\label{subgrad}
\lambda_\ell (t+1) = \mathcal{M}\left\{\lambda_\ell (t) + \beta_t \left(\bar{\|\bm{Q}_\ell\|}_{\pi^\star(\bm{\lambda}(t))} -B_\ell\right)\right\}, \forall \ell \in E,
\end{align}
where $\mathcal{M}(\cdot)$ is the operator that projects the price onto the compact set $[0,K]$ for some sufficiently large $K>0$\footnote{Such a projection is required in order to keep the iterates bounded and non-negative.}, $\bar{\|\bm{Q}_\ell\|}_{\pi^\star(\bm{\lambda}(t))} $ is the steady state average queue lengths at link $\ell$ under the application of $\pi^\star(\bm{\lambda}(t))$, while $\beta_t$ satisfies $\sum_t \beta_t=\infty,\sum_t \beta^2_t<\infty$, and can be taken to be $\frac{1}{t}$. We state the convergence result since it will be utilized later.
\begin{lemma}\label{lemma:pricesubgrad}
Consider the ode
\begin{align}\label{eq:subgradode}
\dot{\bm{\lambda}} = \nabla_{\bm{\lambda}} D(\bm{\lambda}(t)),
\end{align}
where $D(\cdot)$ is the dual function defined in~\eqref{eq:dualprob}. The solution of the ode~\eqref{eq:subgradode} converges to $\bm{\lambda}^\star$, which is the solution of the dual problem~\eqref{eq:dualprob}. Moreover, the price iterations~\eqref{subgrad} track the ode~\eqref{eq:subgradode} in the $\beta_t\to 0,t\to\infty$ limit, and hence the iterations~\eqref{subgrad} converge to $\bm{\lambda}^\star$.   
\end{lemma}
\begin{proof}
In order to analyze the properties of the ode~\eqref{eq:subgradode}, let us consider the evolution of the value of the dual function $D(\bm{\lambda}(t))$. Using the chain rule of differentiation and~\eqref{eq:subgradode}, we have that for a solution $\bm{\lambda}(t)$ of the ode~\eqref{eq:subgradode},
\begin{align*}
\frac{d}{d t}D(\bm{\lambda}(t)) = \|\nabla_{\bm{\lambda}} D(\bm{\lambda}(t))\|^2> 0,
\end{align*}
unless $\nabla_{\bm{\lambda}} D(\bm{\lambda}(t))=0$, i.e., $\bm{\lambda}(t)=\bm{\lambda}^{\star}$.
Since the primal problem, i.e., the CMDP~\eqref{cmdp}-\eqref{cmdp1} is feasible, the dual function is bounded from above. Hence it follows that $\bm{\lambda}(t)\to\bm{\lambda}^{\star}$. 

The connection between the ode and its discrete counterpart, i.e., iterations~\eqref{subgrad} is made using Kushner-Clarke Lemma (see~\cite{borkarbook} Ch: 2 or~\cite{kushner2012stochastic} Ch:2, or~\cite{kushner} Ch:5) according to which, asymptotically the iterates in~\eqref{subgrad} track the ode~\eqref{eq:subgradode}. Hence the $\bm{\lambda}$ iterations~\eqref{subgrad} converge to $\bm{\lambda}^\star$.
\end{proof}
%We note that here the variable $k$ is used to index the price iterations, and must not be confused with the variable $t$, that is used while describing the evolution of network. The price update $\lambda(k)\to \lambda(k+1)$ occurs only when the Q-learning iterations~\eqref{eq:qlearn} with prices set to $\lambda(k)$ have converged.    
\begin{remark}We note that in order to avoid unnecessary notational complexity, we have used $\bm{\lambda}(t)$ to denote the discrete time iterations~\eqref{subgrad}, as well as the continuous time solution of the ode~\eqref{eq:subgradode}. In later sections, we will explicitly mention which of the above two processes is being referred to.
\end{remark}
\subsection{Two Time Scale Stochastic Approximation}\label{subsec:2ode}
As discussed above, successive price updates under the scheme~\eqref{subgrad} need to wait for the Q-learning iterations~\eqref{eq:qlearn} to converge to the policies $\pi^\star_f(\bm{\lambda}(t)),f=1,2,\ldots,F$. Thus, a learning scheme that combines the iterations of Sections~\ref{subsec:ql} and~\ref{subsec:pri} in such a naive fashion, would suffer from slow convergence. An alternative to such a scheme would be to instead perform both the iterations simultaneously, though the price updates be carried out on a slower timescale than that of Q-learning. This would enable the price iterations to view the Q-learning iterations as having converged. Similarly, the Q-learning iterations~\eqref{eq:qlearn} view the link prices $\bm{\lambda}$ as static. Such a scheme would carry out stochastic approximation on two timescales~\cite{borkarscale,borkarbook}. A two timescale stochastic approximation scheme to solve constrained MDPs was proposed in~\cite{borkaractor}. The scheme we introduce now is similar to the one in~\cite{borkaractor}.

The following scheme combines the Q-learning iterations with the price iterations on two different time scales %XX IS EQUILIBRIATED CORRECT WORD XX
\begin{align}
&V(\bm{Q}^f(t),\bm{U}^f(t)) \leftarrow \left\{V(\bm{Q}^f(t),\bm{U}^f(t))\right\} \left(1-\alpha_t\right) \notag\\
&+ \alpha_t \left\{ \sum_{\ell\in R_f}\lambda_\ell Q^f_\ell(t)+\min_{\bm{u}^f}V(\bm{Q}^f(t+1),\bm{u}^f	) - V(\bm{q}_0,\bm{u}_0)\right\}, \label{tts_appendix}\\
&~\qquad \forall f = 1,2,\ldots,F,\notag\\
&\lambda_\ell (t+1) = \mathcal{M}\left\{\lambda_\ell (t) + \beta_t \left(\|\bm{Q}_\ell(t)\|-B_\ell\right)\right\}, \forall \ell \in E,\label{tts1_appendix}
\end{align}
where the sequence $\beta_t$ satisfies $\beta_t=o(\alpha_t)$, i.e., $\lim_{t\to\infty}\frac{\beta_t}{\alpha_t} = 0$. For example, one could let $\alpha_t=\frac{1}{t},\beta_t=\frac{1}{t\left(1+\log t\right)}$. We note that the time scale separation between the price and Q-learning iterations is achieved by allowing the step-sizes $\alpha_t,\beta_t$ associated with their updates to satisfy $\beta_t=o(\alpha_t)$.

The above two-time scale stochastic approximation algorithm can be shown to converge, thereby yielding optimal prices $\bm{\lambda}^\star$, and also the optimal policy $\pi^\star(\bm{\lambda}^\star)$. The analysis of the algorithm~\eqref{tts} is performed using the two timescale ODE method~\cite{kushner,borkarbook,borkarscale} and is similar to the analysis of an actor-critic algorithm for constrained MDPs that was proposed in~\cite{borkaractor}. Hence, we will only provide a proof sketch here.

The following result is Lemma 1 of Ch:6~\cite{borkarbook} (page 66) and will be used in the analysis of multiple timescale stochastic approximation algorithms.
\begin{lemma}\label{lemma:tts}
Consider the following coupled stochastic recursions
\begin{align}
x(t+1) &= x(t) + \alpha_t\left[g(x(t),y(t))+M^{1}(t+1)\right],\label{iterx}\\
y(t+1)&=y(t) + \beta_t\left[h(x(t),y(t))+M^{2}(t+1)\right],\label{itery}
\end{align}
where the iterates $x(t)\in\mathbb{R}^d,y(t)\in\mathbb{R}^k$, and the functions $g:\mathbb{R}^{d+k}\to \mathbb{R}^d,h:\mathbb{R}^{d+k}\to \mathbb{R}^k$, while $\{M^{1}(t)\},\{M^{2}(t)\}$ are bounded martingale difference sequences with respect to the increasing sigma fields $\mathcal{F}_t:=\sigma(x(s),y(s),M^{1}(s),M^2(s),s\leq t),t\geq 0$. Furthermore, the step-sizes $\alpha_t,\beta_t$ satisfy $\sum_t \alpha_t=\infty,\sum_t \beta_t=\infty,\sum_t \alpha_t^2<\infty,\sum_t \beta_t^2<\infty$, and $\frac{\beta_t}{\alpha_t}\to 0$, i.e., the $y(t)$ iterations proceed at a slower timescale than the $x(t)$ iterations. The functions $g,h$ are Lipschitz. Moreover, the ode
\begin{align}\label{eq:ode1}
\dot{x}(t) = g(x(t),y)
\end{align} 
has a globally asymptotically stable equilibrium, denoted $x^\star(y)$. 
Then, for the $(x(t),y(t))$ iterations~\eqref{iterx},\eqref{itery}, we have that 
\begin{align*}
(x(t),y(t)) \to \left\{ \left(x^{\star}(y),y\right):y\in\mathbb{R}^k   \right\}, \mbox{ almost surely },
\end{align*}
i.e., 
$$
\|x(t)-x^{\star}(y(t))\| \to 0,\mbox{ a.s. }.
$$
\end{lemma}
\begin{proof}
See Ch:6 of~\cite{borkarbook}.
\end{proof}
We defined $\bm{V}^{\star}_{\bm{\lambda}}$ to be the vector consisting of optimal Q values when the link prices is set at $\bm{\lambda}$. It follows from Lemma~\ref{eq:bellmanode} that the ode~\eqref{eq:bellmanode} corresponding to Q-learning iterations with prices set to $\bm{\lambda}$ converges to $\bm{V}^{\star}$. Moreover, the price iterations are performed on a slower timescale than Q-learning~\eqref{tts}, i.e., $\frac{\beta_t}{\alpha_t}\to 0$. We can then use Lemma~\ref{lemma:tts} by letting $V(t)$ to be $x(t)$ and $\bm{\lambda}(t)$ to be $y(t)$ in order to infer that asymptotically the routing policy being implemented under the two timescale scheme of~\eqref{tts}-\eqref{tts1} is approximately equal to $\pi^\star(\bm{\lambda}(t))$. That is, if we denote the policy being utilized at time $t$ by the two timescale scheme of~\eqref{tts_appendix}-\eqref{tts1_appendix} by $\pi(t)$, then we have,
\begin{align}\label{eq:inferlemma6}
\|\pi(t)-\pi^\star(\bm{\lambda}(t))\|\to 0,\mbox{ a.s. }.
\end{align}
 Using this result, it can be shown that for the discrete stochastic iterations~\eqref{tts}-\eqref{tts1}, the price vector $\bm{\lambda}(t)$ is ``well approximated" by the following ode,
\begin{align*}
\dot{\lambda}_{\ell} &= \bar{\|\bm{Q}_\ell\|}_{\pi^{\star}(\bm{\lambda})} -B_\ell,\forall\ell\in E,
\end{align*}
or equivalently the ode
\begin{align}\label{eq:priceode:grad}
\dot{\bm{\lambda}} = \nabla_{\bm{\lambda}} D(\bm{\lambda}(t)).
\end{align}
Next, we introduce some machinery in order to make the above statement concrete.

We embed the $\left(\bm{V},\bm{\lambda}\right)$ recursions into a continuous trajectory. Let $s(0)=0$ and for $k=1,2,\ldots$ let $s(k) = \sum_{i=0}^{k}\beta_i$. Now consider the function $\tilde{\bm{\lambda}}$ defined as follows: set $\tilde{\bm{\lambda}}(s(k)) = \bm{\lambda}(k)$, and on each interval of the type $\left[s(k),s(k+1)\right]$ obtain the value of $\tilde{\bm{\lambda}}(t)$ by using linear interpolation on the values $\tilde{\bm{\lambda}}(s(k-1)),\tilde{\bm{\lambda}}(s(k))$. The function $\tilde{\bm{\lambda}}(t)$ thus obtained is continuous and piece-wise linear. Also define $[t]^\prime :=\max\{s(k):s(k)\leq t\}$. For $s>0$, define by $\bm{\lambda}^s(t),t\geq s$, a trajectory of the ode~\eqref{eq:priceode:grad} which satisfies $\bm{\lambda}^s(s) = \tilde{\bm{\lambda}}(s)$. The proof of the following result can be obtained using techniques similar to the proof of Lemma 4.1 of~\cite{borkar2005actor}, and relies on the discussion immediately after Lemma~\ref{lemma:tts}. 
\begin{lemma}\label{eq:price:ode:approx}
For the iterations~\eqref{tts}-\eqref{tts1} we have that 
\begin{align}\label{eq:suffc}
\lim_{s\to \infty}\left(\sup_{t\in \left[s,s+T\right]} \|\tilde{\bm{\lambda}}(t) - \bm{\lambda}^s(t)\|\right) = 0 \mbox{ almost surely}.
\end{align}
\end{lemma}

\begin{proof} [Proof of Theorem~\ref{th:2}]
It follows from Lemma~\ref{lemma:pricesubgrad} that the solution of the ode~\eqref{eq:priceode:grad} converges to the optimal link prices $\bm{\lambda}^\star$. However, from Lemma~\ref{eq:price:ode:approx}, we have that for the \emph{discrete time stochastic} iterations~\eqref{tts}-\eqref{tts1}, the price iterates $\bm{\lambda}(t)$ track the ode~\eqref{eq:priceode:grad} almost surely. Hence, it follows that for the iterations~\eqref{tts}-\eqref{tts1}, we have   $\bm{\lambda}(t)\to\bm{\lambda}^\star$ almost surely. The convergence of the policy $\pi(t)$ being implemented by the two timescale algorithm to the optimal policy $\pi^{\star}(\bm{\lambda}^{\star})$ then follows from~\eqref{eq:inferlemma6}.
\end{proof}

\begin{remark}
Since the underlay links are uncontrollable, the link price iterations can be implemented at the overlay. The overlay however still needs to know the queue lengths $\bm{Q}(t)$ in order to implement the scheme~\eqref{tts}.
\end{remark}
\begin{remark}
Throughout this section, it was assumed that the link-level threshold $\bm{B}$ can be attained under some policy $\pi$. Section~\ref{app:a} discusses the case of infeasible $\bm{B}$.
\end{remark}
\subsection{Analysis of scheme~\eqref{tts}-\eqref{tts1} for the case of infeasible $\bm{B}$}\label{app:a}
If the delay budget $\bm{B}$ is achievable, then it follows from Theorem~\ref{th:2} that the algorithm converges to the optimal policy $\pi^\star$. This section will derive the asymptotic properties of price iteration when $\bm{B}$ is infeasible.

Recall that the price iterations~\eqref{subgrad} were carried out in order to solve the following dual problem 
\begin{align*}
\max_{\bm{\lambda}} D(\bm{\lambda}).
\end{align*}
Since the dual function $D(\bm{\lambda})$ is concave, and the iterates are basically gradient ascent steps, they converge to the optimal price vector $\bm{\lambda}^\star$. However, when the primal problem~\eqref{cmdp} is infeasible, the dual function $D(\bm{\lambda})$, though still concave, is unbounded. An application of the gradient descent method would make the prices $\bm{\lambda}(t)$ unbounded. However since at each time $t$ we project the prices onto a compact set, the price iterations solve the following constrained optimization problem,
\begin{align*}
\max_{\bm{\lambda}: 0\leq  \lambda_\ell\leq K } D(\bm{\lambda}),
\end{align*}
where we note that the quantity $K$ corresponds to the bound on prices $\lambda_\ell$ that is chosen by the scheme~\eqref{tts}-\eqref{tts1}. 
The convergence properties of the algorithm for infeasible $\bm{B}$ then follows by combining~\cite{kushner} Ch: 5, with an analysis similar to that performed in Theorem~\ref{th:2}. The key difference now will be an additional ``forcing term" that appears when the iterates hit the ``upper threshold" value of $K$. This additional forcing term keeps the iterates bounded.   
\begin{lemma}\label{lemma:8}
Consider the underlay network operating under the scheme~\eqref{tts}, or equivalently the algorithm~\ref{alg: end2end} implemented under a fixed value of $\bm{B}$ under which the problem~\eqref{cmdp} is infeasible. The iterates $\bm{\lambda}(t)$ converge to a fixed point of the following ode,
\begin{align*}
\dot{\bm{\lambda}} = \nabla_{\bm{\lambda}} D(\bm{\lambda}) + z(\bm{\lambda}),
\end{align*}
where $z(\bm{\lambda})$ is a ``reflection term" that keeps the prices $\bm{\lambda}(t)$ bounded. 
\end{lemma}
\begin{corollary}\label{coro:1}
Combining Theorem~\ref{th:2} and Lemma~\ref{lemma:8}, we have that when Algorithm~\ref{alg: end2end} is utilized with the delay budget allocation kept fixed at $\bm{B}$, then the prices $\bm{\lambda}(t)\to\bar{\bm{\lambda}}(\bm{B})$.
\end{corollary}
\section{Proof of Lemma~\ref{lemma2}}\label{subsec:replicator}
The proof we provide below is similar to the convergence of the replicator dynamics that is provided in~\cite{borkarbook} Ch:10.
\begin{proof}[Proof of Lemma~\ref{lemma2} ]
Let $S$ denote the $L$ dimensional simplex, i.e.,  
$$S=\left\{\bm{x}: \sum_{i=1}^{L}x_i=1,\quad  x_i\geq 0 \quad \forall i=1,2,\ldots,L\right\}.$$
Let us show that there exists a unique $\frac{\bm{B}^{\star}}{B}\in S$ such that $\frac{\bm{B}^{\star}}{B}$ maximizes $\frac{\bm{B}}{B}\to (\frac{\bm{B}}{B})^{\intercal}\bar{\bm{\lambda}}(\bm{B}^{\star})$. It is easily verified that the set-valued map which maps the vector $\frac{\bm{B}}{B}\in S$ to the set $\{	\bm{x}\in S: \bm{x}\in \arg\max	\bm{x}^\intercal \bar{\bm{\lambda}}(\bm{B}) \}$ is nonempty compact convex and uppersemicontinuous. It then follows from the Kakutani fixed point theorem~\cite{border1989fixed} that there exists a $\frac{\bm{B}^{\star}}{B}\in S$ which maximizes the function $\bm{x}\to \bm{x}^\intercal \bar{\bm{\lambda}}(\bm{B}^{\star})$ over the simplex $S$. Now assume that there exists another $\hat{\bm{B}}\neq \bm{B}^{\star}$, with $\frac{\hat{\bm{B}}}{B}\in S$, such that $\frac{\hat{\bm{B}}}{B}$ maximizes the function $\bm{x}\to \bm{x}^\intercal \bar{\bm{\lambda}}(\hat{\bm{B}})$. This gives us,
\begin{align}
&\left< \frac{\bm{B}^{\star}}{B}-\frac{\hat{\bm{B}}}{B},\bar{\bm{\lambda}}(\bm{B}^{\star})-\bar{\bm{\lambda}}(\hat{\bm{B}}) \right> \notag\\
&= \frac{(\bm{B}^{\star})^\intercal \bar{\bm{\lambda}}(\bm{B}^{\star}) - \left(\bm{B}^{\star}\right)^\intercal \bar{\bm{\lambda}}(\hat{\bm{B}})-(\hat{\bm{B}})^\intercal \bar{\lambda}(\bm{B}^{\star}) + (\hat{\bm{B}})^\intercal\bar{\bm{\lambda}}(\hat{\bm{B}})}{B} \notag\\
&\geq 0,\label{ineq:contradict}
\end{align}
where the inequality follows from our assumptions that $\frac{\bm{B}^{\star}}{B}$ maximizes the function $\frac{\bm{B}}{B}\to (\frac{\bm{B}}{B})^{\intercal}\bar{\bm{\lambda}}(\bm{B}^{\star})$, and
$\frac{\hat{\bm{B}}}{B}$ maximizes the function $\bm{x}^\intercal \bar{\bm{\lambda}}(\hat{\bm{B}})$.  
The inequality~\eqref{ineq:contradict} however contradicts our monotonicity assumption~\eqref{assum1}, and hence we conclude that  $\hat{\bm{B}}=\bm{B}^{\star}$.

Now we show that for the ode~\eqref{delayode},
with $\frac{\bm{B}(0)}{B}$ in the interior of the set $S$, we have that $\bm{B}(t)\to \bm{B}^{\star}$. Consider the Lyapunov function, 
$$\hat{V}(\bm{B}) := \sum_{\ell} B^\star_\ell \ln\left(\frac{ B^\star_\ell}{B_\ell}\right), B\in S.  $$
It follows from Jensen's inequality, that $\hat{V}(\bm{B})\geq 0$, and is $0$ if and only if $\bm{B}=\bm{B}^\star$. Now consider,
\begin{align*}
\frac{\mathrm{d} \hat{V}(\bm{B}(t)) }{\mathrm{d} t}&= -\sum_\ell B^{\star}_\ell \left(\frac{\dot{B}_\ell(t)}{B_\ell (t)}\right)\\
&= \left(	\bm{B}(t)- \bm{B}^{\star}	\right)^\intercal \bar{\bm{\lambda}}(\bm{B}(t))\\
&\leq (\bm{B}(t)- \bm{B}^{\star})^\intercal \bar{\bm{\lambda}}(\bm{B}(t)) - (\bm{B}(t)-\bm{B}^{\star})^\intercal \bar{\bm{\lambda}}(\bm{B}^{\star})\\
&= (\bm{B}(t)- \bm{B}^{\star})^\intercal \left(\bar{\bm{\lambda}}(\bm{B}(t))- \bar{\bm{\lambda}}(\bm{B}^{\star})\right)\\
&<0,
\end{align*}
for $\bm{B}(t)\neq \bm{B}^{\star}$. The first equality follows by substituting for $\dot{B}_\ell(t)$ from~\eqref{delayode} and performing algebraic manipulations, while the first inequality is true because $\bm{B}^{\star}$ maximizes the function $\left(\bm{x}\right)^\intercal \bar{\bm{\lambda}}(\bm{B}^{\star})$. The last inequality follows from the monotonicity assumption~\eqref{assum1}. The claim $\bm{B}(t)\to \bm{B}^{\star}$ follows.
\end{proof}

%XXX NOTE THAT PRICE ITERATIONS AND REPLICATOR DYNAMICS ARE NOT STOCHASTIC...THEY ARE STOCHASTIC ONLY IN THE SENSE THAT ARRIVALS AND CAPACITY ARE STOCHASTIC, AND HENCE NOISE IS AVERAGED OUT AUTOMATICALLY XXXX

\section{Proof of Theorem~\ref{th:4}}\label{subsec:pfth4}
Convergence of multi timescale stochastic approximation techniques is well understood by now, and
the has been covered in detail in Ch:6 of~\cite{borkarbook}. We sketch a proof, and refer the reader to the relevant sections of~\cite{borkarbook} for additional information.

As the first step in our analysis of Algorithm~\ref{alg: end2end}, we would like to show that since the price iterations are performed on a faster timescale that the $\bm{B}(t)$ iterations, the $\bm{B}(t)$ tuner views them as averaged out, and hence the $\bm{\lambda}(t)$ in the equation for $\bm{B}(t)$ update can be approximated by $\bar{\bm{\lambda}}(\bm{B}(t))$.
\begin{lemma}\label{lemma:adhoc2}
For the iterates $\bm{\lambda}(t),\bm{B}(t)$ in Algorithm~\ref{alg: end2end} we have that $(\bm{\lambda}(t), \bm{B}(t))\to \{ (\bar{\bm{\lambda}}(\bm{B}),\bm{B}: \bm{B}\in S )\} $, i.e.,
 $$
\|\bm{\lambda}(t)-\bar{\bm{\lambda}}(\bm{B}(t))\|\to 0
$$
 almost surely, where $\bar{\bm{\lambda}}(\bm{B})$ is the vector of link prices obtained upon convergence of the iterations~\eqref{tts}-\eqref{tts1}.
\end{lemma} 
\begin{proof}
We write down the iterations~\eqref{eq:3ts1}-\eqref{eq:3ts3} on common timescale of $\alpha_t$ as follows
\begin{align}
&V(\bm{Q}^f(t),\bm{U}^f(t)) \leftarrow \left\{V(\bm{Q}^f(t),\bm{U}^f(t))\right\} \left(1-\alpha_t\right) \notag\\
&+ \alpha_t \left\{ \sum_{\ell\in R_f}\lambda_\ell(t) Q^f_\ell(t)\right.\notag\\
&\qquad \qquad\left.\vphantom{\sum_{\ell\in R_f}\lambda_\ell(t) Q^f_\ell(t)}+\min_{\bm{u}^f}V(\bm{Q}^f(t+1),\bm{u}^f	)-V(\bm{q}_0,\bm{u}_0)\right\},\notag\\
&\lambda_\ell (t+1) = \mathcal{M}\left\{\lambda_\ell (t) + \alpha_t \frac{\beta_t}{\alpha_t} \left(\|\bm{Q}_\ell(t)\|-B_\ell(t)\right)\right\},\notag\\
& B^{a}_\ell(t+1)\slash B = B_\ell(t)\slash B\notag\\
&\qquad + \gamma_t \left\{	\frac{B(t)}{B}  \left(\bar{\lambda}_{\ell}(\bm{B} (t)  ) -\sum_{\hat{\ell}} \bar{\lambda}_{\hat{\ell}}(B(t) )\frac{B_{\hat{\ell}} (t)}{B}  \right)	\right\},\notag\\
&\qquad\qquad \forall \ell\in E,t=1,2,\ldots.\notag\\
&\bm{B}(t+1)\slash B = \Gamma\left(\bm{B}^{a}(t)\slash B\right)
\end{align}
Since $\beta_t\slash\alpha_t\to 0,\gamma_t\slash \alpha_t\to 0$, and the processes $\bm{Q}(t)$ and $\bm{B}(t)$ are bounded, the following ode asymptotically approximates the iterates,
\begin{align*}
\dot{\bm{V}} = T\bm{V} - \bm{e} V(\bm{q}_0,\bm{u}_0),\quad
\dot{\bm{\lambda}}(t)  = 0,\quad
\dot{\bm{B}}(t)= 0.
\end{align*} 
where $\bm{e} V(\bm{q}_0,\bm{u}_0)$ is the vector with all entries equal to $V(\bm{q}_0,\bm{u}_0)$.
Thus, following the same arguments that were presented in the discussion immediately after Lemma~\ref{lemma:tts}, we have
\begin{align}\label{eq:conver}
\bm{V}(t) - \bm{V}^{\star}_{\bm{\lambda}(t)} \to 0\mbox{ a.s. },
\end{align}
where $\bm{V}^{\star}_{\bm{\lambda}(t)}$ denotes the optimal Q values when link prices are set to $\bm{\lambda}(t)$.
%\footnote{We note that the convergence $\bm{V}(t) - \bm{V}_{\lambda(t)} \to 0$ does not follow directly from Lemma~\ref{lemma:tts} since now the $V$ iterations also depend upon the $\bm{B}$ iterations, i.e., we are dealing with 3 iterations that are coupled with each other.} 

Next, we re-write evolution of $(\bm{\lambda}(t),\bm{B}(t))$ iterates over multiple time-steps on $\beta_t$ timescale as follows
\begin{align*}
\lambda_\ell(t+k) &= \lambda_\ell(t) + \sum_{s=t}^{t+k}\beta_s \left( \|\bar{\bm{Q}_{\ell}}\|_{\pi^{\star}\left(\bm{\lambda}(s)\right)} - B_\ell(s)\right)\\ 
&+ \Delta(t,t+k),\quad\forall \ell\in E,\\
B_\ell(t+k)\slash B &= B_\ell (t)\slash B  \\
+ \sum_{s=t}^{t+k}\beta_s& \left\{\frac{\gamma_s}{\beta_s} 	\frac{B_\ell(s)}{B}  \left(\lambda_\ell(s) -\sum_{\hat{\ell}} \lambda_{\hat{\ell}}(s)\frac{B_{\hat{\ell}}(s)}{B}\right)	\right\},
\end{align*}
where $\Delta(t,t+k)$ is the error resulting from replacing the term $\|\bm{Q}(s)\|$ by the term $\|\bar{\bm{Q}_{\ell}}\|_{\pi^{\star}\left(\bm{\lambda}(s)\right)} $  during the interval $\{t,t+1,\ldots,t+k\}$\footnote{we have ignored the term being encountered at the ``boundary" that arises due to the projection $\mathcal{M}(\cdot)$. For details see~\cite{kushner}. }.
% \textbf{XX THIS IS NOT CORRECT XX}. 
Using the result~\eqref{eq:conver}, it can be shown that for a fixed $k$, the quantity $\Delta(t,t+k)\to 0$~\eqref{eq:conver} as $t\to\infty$. For a detailed proof, see~\cite{borkar2005actor} or Ch:6 of~\cite{borkarbook}.
Now, since $\gamma_t/\beta_t\to 0$, and the quantity $	B_\ell(t)  \left(\lambda_\ell(t) -\sum_{\hat{\ell}} \lambda_{\hat{\ell}}(t)\frac{B_{\hat{\ell}}(t)}{B}\right)$ is bounded, it follows that the discrete time iterates $\left(\bm{\lambda}(t),\bm{B}(t)\right)$ are asymptotically well-approximated by solution of the following continuous ode,
\begin{align}
\dot{\lambda}_\ell &= \left( \|\bar{\bm{Q}_{\ell}}\|_{\pi^{\star}\left(\bm{\lambda}(t)\right)}  - B_\ell(t) \right),\forall \ell\in E,\\
\dot{\bm{B}}(t)&= 0.
\end{align}
Since $\bar{\bm{\lambda}}(\bm{B})$ is the value of link prices obtained upon convergence of replicator iterations with delay budgets are held fixed at $\bm{B}$, it then follows that for the discrete iterates also, we have $\|\bm{\lambda}(t)-\bar{\bm{\lambda}}(\bm{B}(t))\|\to 0$ \mbox{ a.s. }.
\end{proof}
In view of the above result, we re-write the $\bm{B}(t)$ iterations as follows,
\begin{align}\label{eq:diffdyn}
& \frac{B_\ell(t+1)}{B} - \frac{B_\ell (t)}{B} \notag\\
&= \gamma_t \left\{	B_\ell(t)\slash B  \left(\lambda_\ell(t) -\sum_{\hat{\ell}} \lambda_{\hat{\ell}}(t)B_{\hat{\ell}}(t)\slash B\right)	\right\}\notag\\
&= \gamma_t \left\{	\frac{B_\ell(t)}{B}  \left(\bar{\lambda}_\ell(\bm{B}(t)) -\sum_{\hat{\ell}} \bar{\lambda}_{\hat{\ell}}(\bm{B}(t))\frac{B_{\hat{\ell}}(t)}{B}\right)	\right\} + \gamma_t \delta_1(t),
\end{align}
where, 
\begin{align*}
&\delta_1(t) \\
&= \frac{B_\ell(t)}{B}  \left(\lambda_\ell(t) - \bar{\lambda}_{\ell}(\bm{B}(t)) -\sum_{\hat{\ell}} (\lambda_{\hat{\ell}}(t)-\bar{\lambda}_{\hat{\ell}}(\bm{B}(t)))\frac{B_{\hat{\ell}}(t)}{B}\right)
\end{align*}
is the error term due to approximating the ``true" price $\bm{\lambda}(t)$ by its stationary value $\bar{\bm{\lambda}}(\bm{B}(t))$. Define the piecewise-linear, continuous process $\tilde{\bm{B}}(t)$ in a similar fashion as $\tilde{\bm{\lambda}}(t)$ was defined during the analysis of the price process (similar to the discussion in Section~\ref{subsec:2ode}). We are interested in showing that asymptotically the solutions of the ode~\eqref{delayode} approximate well the values $\bm{B}(t)$ of the discrete recursions in Algorithm~\ref{alg: end2end}. Thus, 
\begin{lemma}\label{lemma:adhoc3}
For a fixed $T>0$, we have that
\begin{align}\label{eq:adhoc1}
\lim_{s\to\infty} \left(\sup_{t\in\left[s,s+T\right]} \|\tilde{\bm{B}}(t) - \bm{B}^s(t)\|\right) = 0 \mbox{ almost surely}.
\end{align} 
\end{lemma}
\begin{proof}
In the below, we let $f_\ell(\bm{B},\bm{\lambda}) = 	\frac{B_\ell}{B} \left(\lambda_\ell(\bm{B}) -\sum_{\hat{\ell}} \lambda_{\hat{\ell}}(\bm{B})\frac{B_{\hat{\ell}}}{B}\right)$.
Consider the following,
\begin{align}\label{eq:adhoc}
&\frac{\tilde{B}_\ell(s(n+m))}{B} = \frac{\tilde{B}_\ell(s(n))}{B} + \int_{s(n)}^{s(n+m)} f_{\ell}(\bm{B}(t),\bar{\bm{\lambda}}(\bm{B}(t))) \mathrm{d}t \notag\\
&+\left(\int_{s(n)}^{s(n+m)} f_{\ell}(\bm{B}([t]^\prime),\bar{\bm{\lambda}}(\bm{B}[t]^{\prime}))- f_{\ell}(\bm{B}(t),\bar{\bm{\lambda}}(\bm{B}(t))) \mathrm{d}t \right)\notag\\
&+ \sum_{k=n}^{n+m}\gamma_k\left\{ f_{\ell}(\bm{B}(k),\bm{\lambda}(k))-f_{\ell}(\bm{B}(k),\bar{\bm{\lambda}}(\bm{B}(k)))\right\},
\end{align}
where $s(n) = \sum_{i=0}^{n-1}\beta_n$. Denote the ``discretization error" in the above by 
\begin{align*}
\Delta_2:=\int_{s(n)}^{s(n+m)} f_{\ell}(\bm{B}([t]^\prime),\bar{\bm{\lambda}}(\bm{B}[t]^{\prime}))- f(\bm{B}(t),\bar{\bm{\lambda}}(\bm{B}(t))) \mathrm{d}t.
\end{align*}
Also let 
\begin{align*}
\Delta_1 =\sum_{k=n}^{n+m}\left(\gamma_k f_{\ell}(\bm{B}(k),\bm{\lambda}(k))-f_{\ell}(\bm{B}(k),\bar{\bm{\lambda}}(\bm{B}(k)))\right).
\end{align*}

Using Gronwall's inequality~\cite{ames1997inequalities}, we have that 
\begin{align*}
\sup_{t\in \left[s,s+T\right]} \|\tilde{\bm{B}}(t) - \bm{B}^s(t)\|\leq K_T\left(\Delta_1+\Delta_2\right),
\end{align*}
where $K_T$ is a suitable constant that depends upon the time interval $T$.

We have $\Delta_2\to 0$ as $n\to\infty$ since the function $f_\ell$ is continuous and the step-sizes $\gamma_t\to 0$.  Also note that the term $\Delta_1$ is equal to $\sum_{k=n}^{n+m}\gamma_k \delta_1(k)$. From Lemma~\ref{lemma:adhoc2}, we have that $\bm{\lambda}(k)\to\bar{\bm{\lambda}}(\bm{B}(k))$. Since the function $f$ is continuous, and $\bm{\lambda}(k)\to\bar{\bm{\lambda}}(\bm{B}(k))$, it follows that $\Delta_2\to 0$.
%This term converges to $0$ as $n\to\infty$ since we have  and $\gamma_t\to 0$.
\end{proof}
\begin{proof}[Proof of Theorem~\ref{th:4}]
The claim follows by combining  Lemma~\ref{lemma:adhoc3} and the fact that the iterates $\bm{B}(t)$ are bounded, with Theorem 2, Ch:2 of~\cite{borkarbook}. %xx explain Kusner Clark result briefly xx 
\end{proof}

\vfil
\begin{IEEEbiography}
    [{\includegraphics[width=1in,height=1.25in,clip,keepaspectratio]{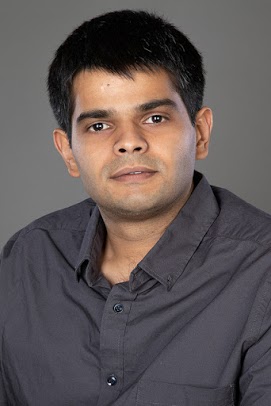}}]{Rahul}
Rahul Singh received the B.E. degree in electrical engineering from Indian Institute of Technology, Kanpur, India, in 2009, the M.Sc. degree in Electrical Engineering from University of Notre Dame, South Bend, IN, in 2011, and the Ph.D. degree in electrical and computer engineering from the Department of Electrical and Computer Engineering Texas A\&M University, College Station, TX, in 2015.

In 2015, he joined as a Postdoctoral Associate at the Laboratory for Information Decision Systems (LIDS), Massachusetts Institute of Technology, where he worked on several stochastic and adversarial decision making problems that arise in network control. He then worked at Encoredtech as a Data Scientist on machine learning problems arising in time-series modeling. He is currently with the Deep Learning Group at Intel. His research interests include decentralized control of large-scale complex cyberphysical systems, operation of electricity markets with renewable energy, machine learning, deep learning, scheduling of stochastic networks serving real time traffic.
\end{IEEEbiography}
\vskip 0pt plus -1fil
\begin{IEEEbiography}
    [{\includegraphics[width=1in,height=1.25in,clip,keepaspectratio]{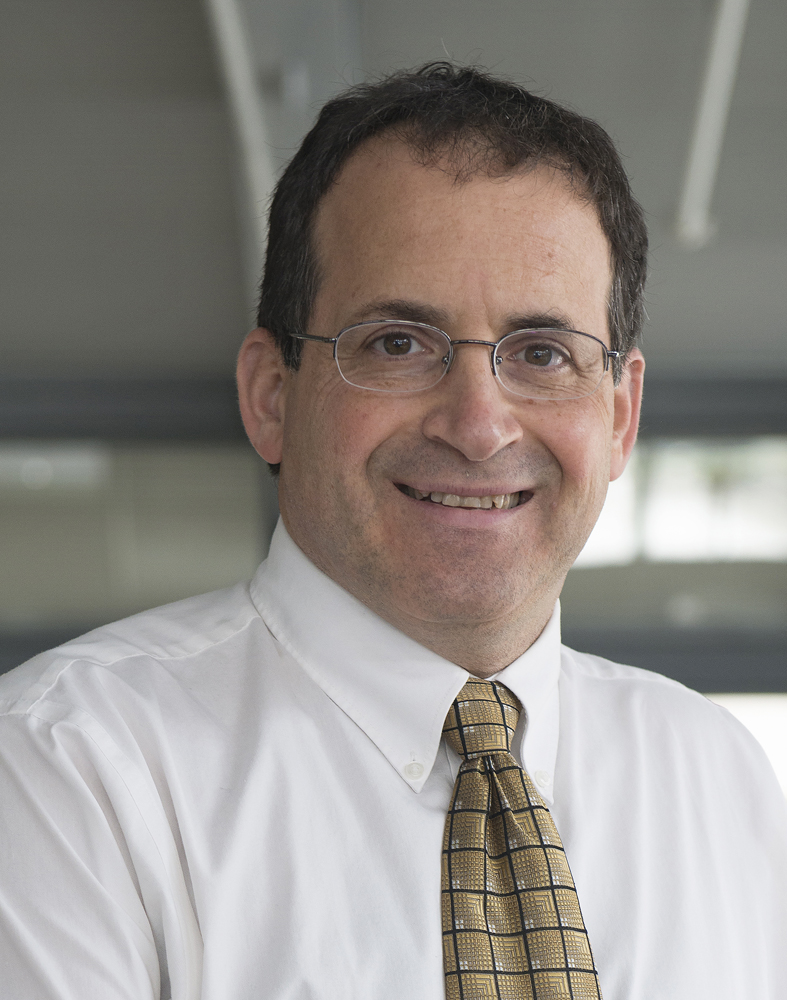}}]{Eytan Modiano}
Eytan Modiano is Professor in the Department of Aeronautics and Astronautics and Associate Director of the Laboratory for Information and Decision Systems (LIDS) at MIT.  Prior to Joining the faculty at MIT in 1999, he was a Naval Research Laboratory Fellow between 1987 and 1992, a National Research Council Post Doctoral Fellow during 1992-1993, and a member of the technical staff at  MIT Lincoln Laboratory between 1993 and 1999.  Eytan Modiano received his B.S. degree in Electrical Engineering and Computer Science from the University of Connecticut at Storrs in 1986 and his M.S. and PhD degrees, both in Electrical Engineering, from the University of Maryland, College Park, MD, in 1989 and 1992 respectively. 

His research is on modeling, analysis and design of communication networks and protocols.   He is the co-recipient of the Infocom 2018 Best paper award, the MobiHoc 2018 best paper award, the MobiHoc 2016 best paper award, the Wiopt 2013 best paper award, and the Sigmetrics 2006 best paper award.  He is the Editor-in-Chief for IEEE/ACM Transactions on Networking, and served as Associate Editor for IEEE Transactions on Information Theory and IEEE/ACM Transactions on Networking.  He was the Technical Program co-chair for  IEEE Wiopt 2006, IEEE Infocom 2007, ACM MobiHoc 2007, and DRCN 2015.  He had served on the IEEE Fellows committee in 2014 and 2015, and is a Fellow of the IEEE and an Associate Fellow of the AIAA.
\end{IEEEbiography}
\end{document}